\documentclass[10pt,journal,compsoc]{IEEEtran}
\long\def\comment#1{}

\ifCLASSOPTIONcompsoc
  \usepackage[nocompress]{cite}
\else
  \usepackage{cite}
\fi
\ifCLASSINFOpdf
\else
\fi
\usepackage{url}
\usepackage{xspace} 
\usepackage{bm}
\usepackage{threeparttable}
\usepackage{stfloats}
\usepackage{algorithm}
\usepackage{algorithmic, amsmath, graphicx}
\usepackage{amsmath,amsthm}
\usepackage{mathrsfs}
\usepackage{booktabs}
\usepackage{subfigure}
\usepackage{color}
\usepackage{rotating}
\usepackage{amsfonts}       % blackboard math symbols
\usepackage{nicefrac}       % compact symbols for 1/2, 
\usepackage{algorithm}
\usepackage{algorithmic}
\usepackage{bm}
\usepackage{threeparttable}
\usepackage{mathrsfs}
\usepackage{lipsum}
\usepackage{amsthm} 
\usepackage{multirow, subfigure}
\usepackage{newfloat}
\usepackage{listings}
\usepackage{wrapfig}

\usepackage[utf8]{inputenc} % allow utf-8 input
\usepackage[T1]{fontenc}    % use 8-bit T1 fonts
\usepackage{hyperref}       % hyperlinks
\usepackage{url}            % simple URL typesetting
\usepackage{booktabs}       % professional-quality tables
\usepackage{amsfonts}       % blackboard math symbols
\usepackage{nicefrac}       % compact symbols for 1/2, etc.
\usepackage{microtype}      % microtypography 
\usepackage{xcolor}         % colors
\usepackage{graphicx} % figures
\usepackage{float} 
\usepackage{wrapfig}
\usepackage{multirow}
\usepackage{multirow} 

\usepackage{amsmath}
\usepackage{amssymb}

\usepackage{amssymb,amsmath}
\usepackage{algorithm}
\usepackage{algorithmic}

\def\ie{$i.e.$}

\usepackage{bm}

\begin{document}
\title{Learning to Accelerate Approximate Methods for Solving Integer Programming via Early Fixing}
% Accelerating Approximate Methods for Solving Integer Programming via Learning to Early Fix 

\author{
Longkang~Li,
        ~Baoyuan~Wu,~\IEEEmembership{Member,~IEEE}
\IEEEcompsocitemizethanks{\IEEEcompsocthanksitem Longkang Li and Baoyuan Wu are with the School of Data Science, The Chinese University of Hong Kong, Shenzhen, China and also with Secure Computing Lab of Big Data, Shenzhen Research Institute of Big Data, Shenzhen, China. \protect\\
%E-mail: \{lilongkang, wubaoyuan\}@cuhk.edu.cn.
\IEEEcompsocthanksitem
Correspondence to Baoyuan Wu (wubaoyuan@cuhk.edu.cn).
% \IEEEcompsocthanksitem J. Doe and J. Doe are with Anonymous University.
}% <-this % stops an unwanted space

% \thanks{Manuscript under review.}

}


\markboth{}%
% \markboth{Manuscript for IEEE Transactions on Pattern Analysis and Machine Intelligence}%
{Shell \MakeLowercase{\textit{et al.}}: Bare Demo of IEEEtran.cls for IEEE Journals}
\IEEEtitleabstractindextext{%
\begin{abstract}
Integer programming (IP) is an important but challenging problem. Approximate methods have shown promising performance on solving the IP problem. However, we observed that a large fraction of variables solved by some iterative approximate methods fluctuate around their final converged discrete states in very long iterations. 
It implies that these approximate methods could be significantly accelerated by early fixing these fluctuated variables to their converged states, while not significantly harming the solution quality. To this end, we propose an innovative framework of learning to early fix variables along with the approximate method. Specifically, we formulate the early fixing process as a Markov decision process, and train it using imitation learning, where a policy network evaluates the posterior probability of each free variable concerning its discrete candidate states in each block of iterations. 
Extensive experiments on three typical IP applications are conducted, including constrained linear programming, MRF energy minimization and sparse adversarial attack, covering moderate and large-scale IP problems. 
The results demonstrate that our method could not only significantly accelerate the previous approximate method up to over 10 times in most cases, but also produce similar or even better solutions.
The implementation of our method is publicly available at  \url{https://github.com/SCLBD/Accelerated-Lpbox-ADMM}.

\end{abstract}

}

\maketitle

\IEEEdisplaynontitleabstractindextext

% \fi
\IEEEpeerreviewmaketitle

\IEEEraisesectionheading{\section{Introduction}\label{sec:introduction}}

\IEEEPARstart{I}{nteger} programming (IP) is an important and challenging problem in many fields, such as machine learning \cite{khalil2016machine} and computer vision \cite{wang2014tracking}. IP can be a versatile modeling tool for discrete or combinatorial optimization problems with a wide variety of applications, and thus has attracted considerable interests from the theory and algorithm design communities over the years \cite{gunluk2021optimal}. There are rich literature and a wide range of developed methods and theories for solving IP. Generally, we could divide them into two categories: exact methods and approximate methods. Some exact methods are widely utilized, such as branch-and-bound \cite{lawler1966branch} , cutting plane \cite{kelley1960cutting} and branch-and-cut \cite{mitchell2010branch} methods. The branch-and-bound method \cite{lawler1966branch} is an approach that partitions the feasible solution space into smaller subsets of solutions. The cutting-plane method \cite{kelley1960cutting} is any of a variety of optimization methods that iteratively refine a feasible set or objective function by means of linear inequalities, termed \textit{cuts}. The branch-and-cut \cite{mitchell2010branch} method combines branch-and-bound and the cutting plane method. These exact methods are able to get the optimal solutions, however, they are usually suffering from time-consuming issues due to the repeated solving of relaxed linear problems. Thereafter, in the recent years, more and more research focuses on the approximate methods, where a feasible solution is obtained within the limited time. Linear relaxation\cite{boyd2004convex} relaxes the binary constraints $x \in \{0,1\}$ to the box constraints $x \in [0,1]$. Spectral relaxation\cite{zha2001spectral} relaxes the binary constraint to the $\ell_2$-ball, leading to a non-convex constraint. As regard to the SDP relaxation\cite{lasserre2001explicit}, the binary constraints are substituted with a positive semi-definite matrix constraint, \ie, $\textbf{X} \succcurlyeq 0$.   

\begin{figure*}[!t] %H为当前位置，!htb为忽略美学标准，htbp为浮动图形
  \centering %图片居中
  \includegraphics[width=0.99\textwidth]{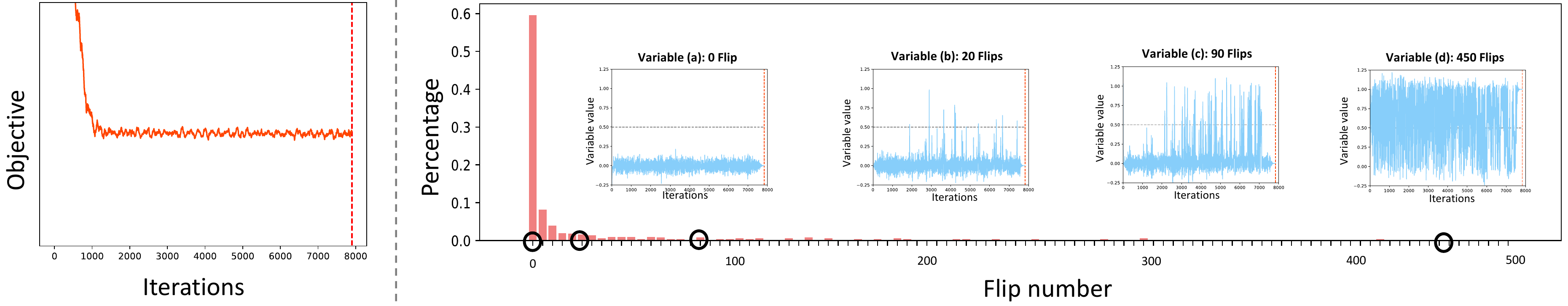}
  \caption{A constrained linear programming instance with 500 variables is solved by $\ell_p$-box ADMM, converging after 7827 iterations. \textbf{\textit{Left}}: We record how the objective changes with respect to the iterations. \textbf{\textit{Right}}: a flip histogram. We use "Flip number" to evaluate the iterating stableness of the variable. For one variable, if the values of two adjacent iterations go across 0.5, we call it a 'Flip'. When converged, each variable gets its Flip number. We build the percentage histogram of all these 500 variables according to their Flip number, where the minimum is 0, the maximum is 450 Flips, and the horizontal axis has 5 Flips as a interval. The results show that 59.6\% of variables have [0,5) Flips, among which 34.0\% have 0 Flip. We also present 4 different variable iterating processes, corresponding to 0, 20, 90, 450 Flip(s).}
  \label{fig0} %用于文内引用的标签
\end{figure*}

Besides the above mentioned relaxation-based approximate methods for integer programming, recently there has been increasing attention to another type of approximate methods, which is iterative and based on the alternating direction method of multipliers (ADMM) \cite{boyd2011distributed}. ADMM is a powerful algorithm for distributed convex optimization, with an attempt to blend the benefits of dual decomposition \cite{rantzer2009dynamic} and augmented Lagrangian methods \cite{chatzipanagiotis2015augmented}. ADMM coordinates the solutions of small local subproblems to find a solution of the large global problem. Many variants of ADMM have been proposed with different purposes of better acceleration, convergence, or stability, and have been applied to solving different types of optimization tasks. Bethe ADMM \cite{fu2013bethe} was proposed for tree decomposition based parallel MAP inference, which used an inexact ADMM augmented with a Bethe-divergence based proximal function. Bregman ADMM \cite{wang2013bregman} was then proposed, which provided a unified framework for ADMM. Bregman ADMM then has a number of variants such as generalized ADMM and inexact ADMM. Linearized ADMM \cite{xie2019differentiable} \cite{liu2019linearized} was also proposed for convex optimization. One state-of-the-art ADMM method for solving IP is $\ell_p$-Box ADMM \cite{wu2019lp}, where the binary constraints are equivalently replaced by the intersection of a box constraint and a $\ell_p$-norm sphere constraint.

Towards those ADMM based approximate methods, we observed that regarding the approximate methods a large fraction of variables fluctuated around theirs final converged states in very long iterations. In Fig. \ref{fig0}, we solve a constrained linear programming instance with 500 variables by $\ell_p$-box ADMM, which converges after 7827 iterations. The left figure illustrates the objective changes with respect to the iterations. And we can see that the convergence reaches after a long fluctuation. In the right figure, We introduce "Flip number" to evaluate the iterating stableness of the variable. For one variable, if the values of two adjacent iterations go across 0.5, we call it a 'Flip'. For example, the the variable value at $t$-th iteration is 0.9 and that at $t+1$-th iteration is 0.3, thus it is a Flip. Each variable corresponds to one Flip number, the smaller the Flip number, the more stable the variable iterating. According to the Flip histogram, 59.6\% of variables have [0,5) Flips, among which 34\% have 0 Flip. Most variables has small Flip numbers. To that end, we believe that a large proportion of variables are fluctuating around their final converged states (0 or 1) within small ranges. Currently, one solution for algorithmic acceleration is to early stop the iterations \cite{zamir2017feedback} \cite{huang2017multi}. However, early stopping has two shortcomings: 1). There is trade-off between the objective efficiency and the runtime effectiveness, when stopping much earlier the objective accuracy may decrease more. 2). It always use the whole set of variables to consider whether and when to stop. 

\begin{figure*}[!t] %H为当前位置，!htb为忽略美学标准，htbp为浮动图形
  \centering %图片居中
  \includegraphics[width=1\textwidth]{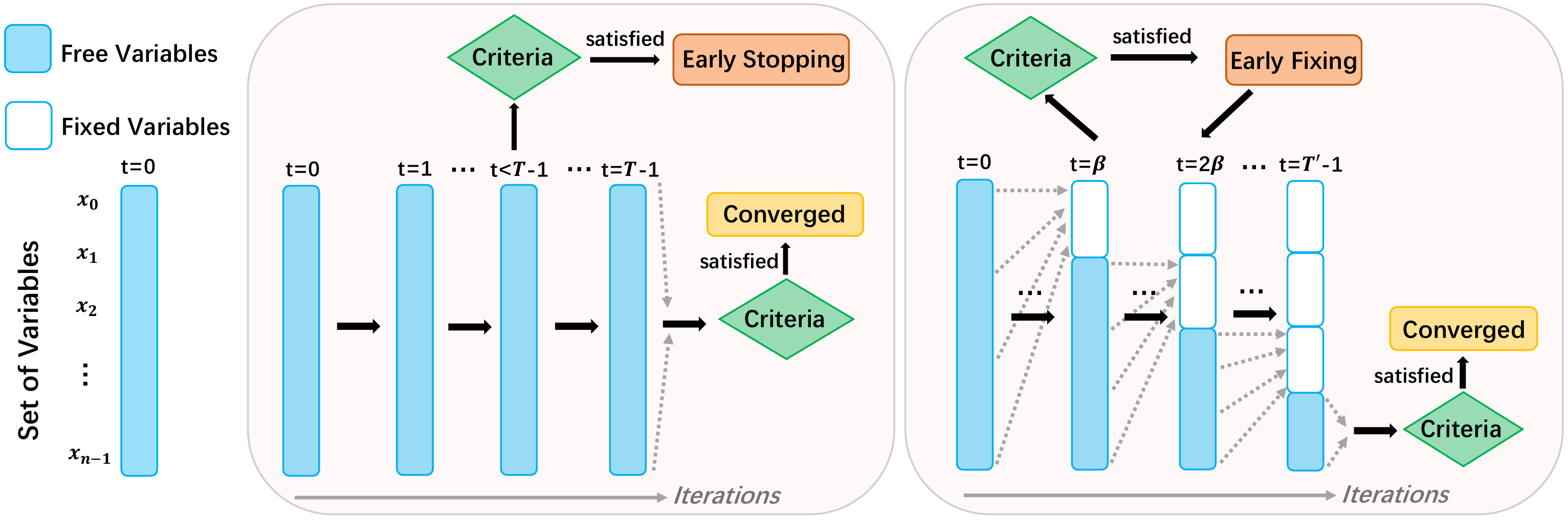} %插入图片，[]中设置图片大小，{}中是图片文件名
  \caption{Comparison between Early Stopping \textit{(Left)} and Early Fixing \textit{(Right)}: 1) Early stopping only considers the depth (iteration steps), while early fixing also thinks over the width (variable dimensions). 2) Early stopping regards the set of variables as a whole, while early fixing treats every single variable independently. 3) Decisions on whether to early stop are made in every iteration, while those for early fixing are once every block of iterations.} %最终文档中希望显示的图片标题
  \label{fig1} %用于文内引用的标签
\end{figure*}

Inspired by this observation, we were thinking: \textit{why not take every single variable independently and then asynchronously fix them instead of early stopping all of the variables at one iteration?} To the end, we propose an early fixing framework, which aims to accelerate the approximate method by early fixing these fluctuated variables to their converged states while not significantly harming the converged performance. Fig. \ref{fig1} shows the comparison between early stopping and early fixing. And there are mainly three differences: \textit{Firstly}, early stopping does only consider the depth of optimization, \ie, the number of iterations, while early fixing also thinks over the width of optimization, \ie, the dimension of variables. \textit{Secondly}, early stopping regards the set of variables as a whole, while early fixing treats every single variable independently. And \textit{thirdly}, decisions on whether to early stop are made in every single iteration, while those for early fixing are once every block of iterations, \ie, $\beta$ iterations.

Under our proposed early fixing framework, in each block of iterations, given the iterative values of the variables within the past $\beta$ iterations, a policy network will evaluate the posterior probability of each variable concerning all discrete candidate states. If the posterior probability with respect to one state exceeds the fixing threshold, then the action of fixing this variable to that discrete state will be conducted, and this variable will not be updated in later iterations; otherwise, no fixing action will be conducted and this variable will be further updated. Specifically, for each variable, the continuously iterative values within the past $\beta$ iterations are sequential according to the time series. Recently, the Transformer structure \cite{vaswani2017attention} has exhibited powerful performance in the sequential networks though the multi-headed attention (MHA) mechanism. Herein, we incorporate the attention layers in our policy network. When solving a problem with early fixing framework, one block of iterations only decides a proportion of variables to conduct early fixing, thus the process is episodic until termination. We can regard the solving process as a Markov decision process \cite{howard1960dynamic} and train it using imitation learning \cite{torabi2018behavioral}. Since the input of policy network only requires the iterative values of variables, without any other constraint information, thus our early fixing framework can be versatile enough, available to all the IP problems of any orders or types, no matter linear or quadratic, constrained or unconstrained. 

In this paper, in order to accelerate the ADMM based approximate methods for solving the IP problems, especially improving the scalability of the IP problems, we propose the early fixing framework combined with learning techniques. The main contributions of this paper are four-fold:

\begin{itemize} 

\item[(i)] To the best of our knowledge, we are the first to propose an \textit{early fixing} framework to accelerate the approximate methods, where the variables are treated independently with one another and we can asynchronously fix them according to their continuously iterative values within the past series of iterations. Once fixed, the variables will not be further updated. And free variables will continue iterating and updating.

\item[(ii)] We formulate the whole early fixing process when accelerating solving an IP problem as a Markov decision process, and train it using behaviour cloning as a method of imitation learning. We also incorporate the weighted binary cross-entropy (WBCE) loss during the training.

\item[(iii)] We adopt the learning techniques with the attention layers in our policy network, to decide whether to fix the variable or not. 
% to obtain better feature representations.
% Rather than the hand-designed functions, w
 
\item[(iv)] We apply our proposed early fixing framework to three different IP applications: constrained linear programming, MRF energy minimization and sparse adversarial attack. The former one is linear IP problem, while the latter two are quadratic IP problems. We extend the problem scale from regular size to significantly large size. The extensive experiments reveal the competitiveness of our early fixing framework: the runtime speeds up significantly, while the solution quality does not degrade much, even in some cases it is available to obtain better solutions. 

\end{itemize}

The rest of this paper is organized as follows. Section \ref{rw} briefly reviews the related work. Some background information is given in Section \ref{background}. The details of our early fixing framework are described in Section \ref{method}. Section \ref{linear}-\ref{saa} show the extensive applications of our proposed framework in the areas of constrained linear programming, MRF energy minimization and sparse adversarial attack, respectively. We present the conclusion in Section \ref{conclusion}. 

This is the testing $\bm{a}$ \textbf{a} $\delta$ $\in$ $\geq$. $\mathbb{R}$

\section{Related work}
\label{rw}

\textbf{Integer programming} \ There are rich literature and a wide range of developed methods and theories for solving IP. Generally, we could divide them into two categories: exact methods and approximate methods. Some exact methods are widely utilized, such as branch-and-bound \cite{lawler1966branch} , cutting plane \cite{kelley1960cutting} and branch-and-cut \cite{mitchell2010branch} methods. The branch-and-bound method \cite{lawler1966branch} is an approach that partitions the feasible solution space into smaller subsets of solutions. The cutting-plane method \cite{kelley1960cutting} is any of a variety of optimization methods that iteratively refine a feasible set or objective function by means of linear inequalities, termed \textit{cuts}. The branch-and-cut \cite{mitchell2010branch} method combines branch-and-bound and the cutting plane method. These exact methods are able to get the optimal solutions, however, they are much too slow in the runtime due to the repeated solving of relaxed linear problems. In order to obtain feasible solutions within the given time, some approximate methods are proposed. Linear relaxation\cite{boyd2004convex} relaxes the binary constraints $x \in \{0,1\}$ to the box constraints $x \in [0,1]$. Spectral relaxation\cite{zha2001spectral} relaxes the binary constraint to the $\ell_2$-ball, leading to a non-convex constraint. As regard to the SDP relaxation\cite{lasserre2001explicit}, the binary constraints are substituted with a positive semi-definite matrix constraint, \ie, $\textbf{x} \succcurlyeq 0$.

\noindent\textbf{Integer programming by ADMM} \ Besides the above mentioned relaxation-based approximate methods for integer programming, recently there has been increasing attention to another type of approximate methods, which is iterative and based on the alternating direction method of multipliers (ADMM) \cite{boyd2011distributed}. ADMM is a powerful algorithm for distributed convex optimization. With an attempt to blend the benefits of dual decomposition \cite{rantzer2009dynamic} and augmented Lagrangian methods \cite{chatzipanagiotis2015augmented}, ADMM coordinates the solutions of small local subproblems to find a solution of the large global problem. Many variants of ADMM have been proposed with different purposes of better acceleration, convergence, or stability, and have been applied to solving different types of optimization tasks. Bethe ADMM \cite{fu2013bethe} was proposed for tree decomposition based parallel MAP inference, which used an inexact ADMM augmented with a Bethe-divergence based proximal function. Bregman ADMM \cite{wang2013bregman} was then proposed, which provided a unified framework for ADMM and its variants, including generalized ADMM, inexact ADMM and Bethe ADMM. Linearized ADMM \cite{xie2019differentiable} \cite{liu2019linearized} was also proposed for convex optimization. One state-of-the-art ADMM method for solving IP is $\ell_p$-box ADMM \cite{wu2019lp}, where the binary constraints are equivalently replaced by the intersection of a box constraint $S_b$ and a $\ell_p$-norm sphere constraint $S_p$. The equivalent replacement can be formulated as: $\textbf{x} \in \{0,1\}^n \  \Leftrightarrow \  \textbf{x} \in S_b \cap \textbf{x} \in S_p \Leftrightarrow \  \textbf{x} \in [0,1]^n \cap \{\textbf{x}: \lVert \textbf{x} - \frac{1}{2}\textbf{1} \rVert ^p _p = \frac{n}{2^p} \}$, then additional variables are introduced to separate the constraints, so as to coordinate with ADMM method. This method could give a quasi-optimal solution at the globally converged point. $\ell_p$-box ADMM has wide application in MAP inference \cite{wu2020map}, sparse adversarial attack \cite{fan2020sparse}, model compression \cite{li2019compressing}, feature selection \cite{zhang2020top} and etc. Though $\ell_p$-Box ADMM builds the bridge between IP and continuous optimization algorithms and achieves great performances in solving IP tasks, it still remains the issues of scalability. 

\begin{table*}[!t]
    \centering
\caption{Summary of notations} 
\label{ablation}
\resizebox{17.9cm}{!}{
\begin{tabular}{cl|cl}
    \toprule[1.5pt]
     {Notation} & {Meaning} & Notation & Meaning \\
    \midrule
    $n$ & number of variables, $n$>0. & $m$ & number of constraints, $m\geq$0. \\
    $t$ & iteration index, $t \in \{0,...,\textit{T}{-}1\}$. & $i$ & variable index, $i \in \{0,...,n{-}1\}$. \\
    $T$ & maximum iteration without early fixing, $T>0$. & $T'$ & maximum iteration with early fixing, $T'>0$. \\
    $\delta$ & fixing threshold, deciding whether to fix, $\delta\in$[0.5,1]. & $\beta$ & block size, denoting one block of iterations, $\beta$>1. \\
    \midrule 
    $f(\cdot)$ & objective function, linear or quadratic. & $\mathcal{C}$ & set of constraints, $\mathcal{C} \in  \mathbb{R}^m$. \\ 
    $\textbf{x}$ & set of variables, $\textbf{x}$ $\in$ $\mathbb{R}^n$. & $\textbf{A}$ & matrix in objective function, $\textbf{A} \in \mathbb{R}^{n \times n}$. \\
    $\textbf{b}$ & vector in objective function, $\textbf{b}$ $\in \mathbb{R}^n$. & $\textbf{C}$ & matrix in constraint set, $\textbf{C} \in \mathbb{R}^{m \times n}$. \\ 
    $\textbf{d}$ & vector in constraint set, $\textbf{d}$ $\in \mathbb{R}^m$. & $\bigotimes$ & any relational symbol. <, >, =, $\geq$ or $\leq$. \\ 
    \midrule
     $\pi(\cdot)$ & policy network for early fixing. & $\theta$ & weights of policy network. \\
     $y, \bm{y}$ & $\bm{y}$ is the iterative values of variable $y$. $\bm{y}\in \mathbb{R}^{\beta\times1}$ & $\bm{z}$ & iteration embedding of one variable, $\bm{{z}}\in \mathbb{R}^{\alpha\times{d_h}}$. \\
     $\alpha$ & node number for iteration embedding, $\alpha>1$. & $\bm{\hat{z}}$ & iteration embedding with positional encoding, $\bm{\hat{{z}}}\in \mathbb{R}^{\alpha\times{2d_h}}$. \\
     $d_h$ & node dimension, $d_h{\geq}1$. & $d_n$ & hidden dimension, $d_n$=128. \\ 
     \midrule
     $k$ & position index, $k \in \{1,...,\alpha\}$. & $j$ & dimension index for positional encoding, $2j\leq d_h$. \\
     $\bm{\hat{z}}_k$ & node input in attention layers, $\bm{\hat{z}}_k \in \mathbb{R}^{2d_h}$ & $\bm{h}_k, \bm{\hat{h}}_k$ & node embeddings in attention layers. $\bm{h}_k, \bm{\hat{h}}_k \in \mathbb{R}^{d_n}$. \\
     $\mathbb{L}$ & number of layers, $\mathbb{L}$>1. & $\ell$ & layer index in attention layers, $\ell \in \{1,...,\mathbb{L}\}$.\\
     $H$ & number of heads in MHA sublayer, $H=$8. & $d_{n'}$ & hidden dimensions in FF sublayer, $d_{n'}=$512.\\
     \midrule
    $\bm{\overline{z}}$ & concatenated embedding, $\bm{\overline{z}} \in \mathbb{R}^{(\alpha{\cdot}{d_n}) {\times} 1}$. & $\bm{p}$ & the probability vector, each element $p_i\in[0,1]$. \\
    $u$ & number of free variables, $u>0, u\leq n$. & $v$ & number of fixed variables, $v>0, v\leq n$. \\ 
    $r$ & number of rounds for conducting early fixing, $r>1$. & $\mathcal{M}$ & the approximate method to be accelerated.\\
    $\gamma$ & number of blocks, used in the network training, $\gamma$>1. & N & number of training instances, N>1.\\
    \midrule 
    e & instance index, $e \in \{0,...,\textit{N}{-}1\}$. & $q$ & one element of loss, a scalar. \\
    $w$ & weight for training, a scalar. & $\mathcal{J(\cdot)}$ & the binary cross entropy loss for network training. \\ 
    $\mathcal{I}$ & one instance as Formulation (\ref{eq1}). & N' & number of instances for inference, N'>1. \\
    $\xi$ & a constant for linear programming.  & $\bm{\epsilon}$ & perturbation for sparse adversarial attack. \\
    $\bm{\zeta}$ & the vector of perturbation magnitudes. & $\bm{\eta}$ & the vector of perturbed positions. \\ 
    \bottomrule[1.5pt]
\end{tabular}}
\end{table*}

\noindent\textbf{Integer programming meets learning} \ The application of machine learning (ML) to discrete optimization has been a popular topic with various approaches in the literature \cite{bengio2020machine}. Learning to branch is an interesting topic attracting a great deal of attention. Khalil et al. \cite{khalil2016learning} took the first step towards statistical learning of branching rules in BB. Alvarez et al. \cite{alvarez2017machine} and Gasse et al. \cite{gasse2019exact} learn a branching rule offline on a collection of similar instances, and the branching policy is learned by the imitation of the strong branching expert. Graph Neural Network (GNN) approach for learning to branch has successfully reduced the runtime \cite{gasse2019exact}. Gupta et al. \cite{gupta2020hybrid} consider the availability of expensive GPU resources for training and inference, thus devise an alternate computationally inexpensive CPU-based model that retains the predictive power of the GNN architecture. Tang et al. \cite{tang2020reinforcement} utilize reinforcement learning for learning to cut. Recently there is one set of approaches focusing on directly learning the mapping from an IP to its approximate optimal solution, instead of solving the IP by any exact or approximate solvers. Vinyals et al. \cite{vinyals2015pointer} introduce the pointer network as a model that uses attention to output a permutation of the input, and train this model offline to solve the TSP problem. Bello et al. \cite{bello2016neural} introduce an Actor-Critic algorithm to train the pointer network without supervised solutions. Kool et al. \cite{kool2018attention} propose a model based on attention layers \cite{vaswani2017attention} to solve the routing problems. Andrychowicz et al. \cite{andrychowicz2016learning} and Li et al. \cite{li2016learning} propose learning to optimize or learning to learn, casting an optimization problem as a learning problem. Nowak et al. \cite{nowak2017note} train a Graph Neural Network in a supervised manner to directly output a tour as an adjacency matrix, which is converted into a feasible solution by a beam search. 

Most of the above mentioned papers use the networks to entirely substitute with the optimizer, different from that, in this paper we simply utilize the network to accelerate the optimization, just like an attachment module. 

\noindent\textbf{Imitation learning} Imitation learning (IL) techniques aim to mimic the hebaviour from an expert or teacher in a given task \cite{hussein2017imitation}. IL and reinforcement learning (RL) both work for the Markov decision processes (MDP). RL tends to have the agent learn from scratch through its exploration with a specified reward function, however, the agent of IL does not receive task reward but learn by observing and mimicing \cite{torabi2019recent}. Similar to traditional supervised learning (SL) where the samples represent pairs of features and ground-truth labels, IL has the samples demonstrating pairs of states and actions. One fundamental difference between SL and IL is that: SL follows the assumption that the training and test data are independent and identically distributed (IID), while those of IL are Non-IID where the current state is only correlated to the previous state. Broadly speaking, research in the IL can be split into two main categories: behavioral cloning (BC) \cite{torabi2018behavioral}, and inverse reinforcement learning (IRL) \cite{abbeel2004apprenticeship}. In this paper, we choose BC for the training. 

\noindent\textbf{Early exiting and early stopping} \ The term "early exiting" originally comes from computer vision and image recognition, which is mainly aimed at improving the computation efficiency based on specific architectures during the inference phase \cite{zamir2017feedback} \cite{huang2017multi}. As networks continue to get deeper and larger, these costs become more prohibitive for real-time applications. To address the issue, the proposed architecture exits the network early via additional branch classifiers with high confidence \cite{teerapittayanon2016branchynet}. Early stopping is a form of regularization used to avoid overfitting when training a learner with an iterative method, such as gradient descent \cite{prechelt1998early, yao2007early}. When a certain criterion is satisfied, early stopping will be conducted before the ultimate convergence. Kaya et al. \cite{kaya2019shallow} proposes to avoiding “over-thinking” by early stopping, where the deep neural networks can reach correct predictions before the final layers to save the running time. In optimal control literature, optimal stopping \cite{shiryaev2007optimal} is a problem of choosing a time to take a given action based on sequentially observed random variables in order to maximize an expected payoff. Optimal stopping can be seen as a special case of early stopping. Becker et al. \cite{becker2019deep} and Chen et al. \cite{chen2020learning} use deep reinforcement learning to learn the optimal stopping policy. Besides, Chen et al. \cite{chen2020learning} provide a variational Bayes perspective to combine learning to predict with learning to stop. 
In this paper, we propose a novel early fixing framework for accelerating solving the generic IP problems. Whether early exiting or early stopping, the models focus on the depth of iterations, while our proposed early fixing also considers the width of variable dimensions. 

% The process will be early terminated before reaching the final converged states when the given criterion is satisfied. Different from that, our proposed early fixing framework does not only consider the depth of iterations, but also take the width of the variable dimensions into account. Figure \ref{fig2} explicitly demonstrates the comparison between early stopping and early fixing.

% From the perspective of width, early stopping regards the set of variables as a whole, while early fixing treats every single variable independently to do evaluations. Moreover, from the perspective of depth, Decisions on whether to early stop are made in every single iteration, while those for early fixing are once every block of iterations. 

% \noindent\textbf{Fix-and-optimize} \ Fix-and-optimize \cite{voss2009matheuristics} is a metaheuristic for solving mixed integer linear programming (MIP) that iteratively decomposes a problem into smaller subproblems. The solution obtained in each iteration becomes the current solution when it improves the objective value. In further iterations of the algorithm, a different group of variables is selected to be optimized. 

\noindent\textbf{Fix-and-optimize} \ Fix-and-optimize \cite{gintner2005solving} is a metaheuristic, firstly proposed by Gintner et. al. for solving mixed integer linear programming (MIP), which iteratively decomposes a problem into smaller subproblems. In each iteration, a decomposition process is applied aiming at fixing most of the decision variables at their value in the current solution. Since the resulting subproblem is composed only by a small group of free variables to be optimized, each subproblem can be solved fairly quickly by a MIP solver, when compared with the full model. The solution obtained in each iteration becomes the current solution when it improves the objective value. In further iterations, a different group of variables is selected to be optimized. This process is repeated until a termination condition is satisfied. The fix-and-optimize algorithm has wide applications in lot sizing problem \cite{helber2010fix}, timetabling problem \cite{dorneles2014fix}, and etc. Different from the fix-and-optimize algorithm where the fixed variables in previous iteration will be released in the next iteration, our early fixing framework requires that the fixed variables will stay fixed and not appear in the following iterations.

\noindent\textbf{Variable fixing} \ The strategy of fixing variables \cite{wang2011effective} within optimization algorithms often proves useful for enhancing the performance of methods for solving constraint satisfaction and optimization problems. Such a strategy has come to be one of the basic strategies associated with Tabu search \cite{gendreau2005tabu}. Two of the most important features in Tabu search are how to score the variables (variable scoring) and which variables should be fixed (variable fixing). Similar to fix-and-optimize algorithm, the fixed variables in Tabu search could possibly be freed, while those in our early fixing framework will keep fixed and not released in the next iterations.

\section{Background}
\label{background}

\subsection{Problem definition} 
Throughout this paper, we focus on the problem of generic IP problems, which can be generally formulated as a binary mathematical optimization problem as follows:

% \vspace{-2ex}
\begin{align}
    \label{eq1}
   \mathop{\arg\max}_{\textbf{x}} \ f(\textbf{x}), \ \  s.t. \ \ \textbf{x} {\in} \mathcal{C}, \ \ \textbf{x}{\in}\{0,1\}^n,
\end{align}

\noindent where $\mathcal{C} \in  \mathbb{R}^m$ is the set of constraints, and $\textbf{x} \in \mathbb{R} ^n$ is the set of binary variables. $n, m$ denotes the number of variables and constraints, respectively. In this paper, we mainly focus on the linear and quadratic IP problems. 

\subsection{Approximate methods} 
The approximate methods for solving IP problems are based on a bunch of iterations, which is a mathematical procedure that uses an initial value to generate a sequence of improving approximate solutions until the convergence. In recent year, there have been a great deal of works utilizing ADMM to solve the IP problems \cite{wang2013bregman, xie2019differentiable, liu2019linearized, wu2019lp, wu2020map, zhang2020top}. Therein, in this paper we propose the early fixing framework, aiming to accelerate these ADMM-based approximate methods.

\section{Early Fixing Framework}\label{method}
In this section, we give an overview of how our proposed early fixing framework is utilized to accelerate the iterative approximate methods for solving generic IP problems. We firstly formulate the early fixing process as a Markov decision process (MDP), as described in subsection \ref{mdp}. We present the overview of the policy network in subsection \ref{learn2fix}. We then present how to train the policy network with the weighted binary cross-entropy loss in a imitation learning method in subsection \ref{training}. Last but not least, we exhibit how to do the inference with our early fixing framework in subsection \ref{updating}, some mathematical propositions and proofs are given as regard to the problem reformulations.

% As a matter of fact, there are two important modules within the early fixing framework: how to decide for each variable when and whether to fix or not, and how to update the IP problems after some variables being fixed. We will introduce how to fix in the subsection \ref{learn2fix} with the attention policy network and how to update in the subsection \ref{updating} with mathematical proposition and proof. 

\begin{figure}[!t] %H为当前位置，!htb为忽略美学标准，htbp为浮动图形
  \centering %图片居中
  \includegraphics[width=0.40\textwidth]{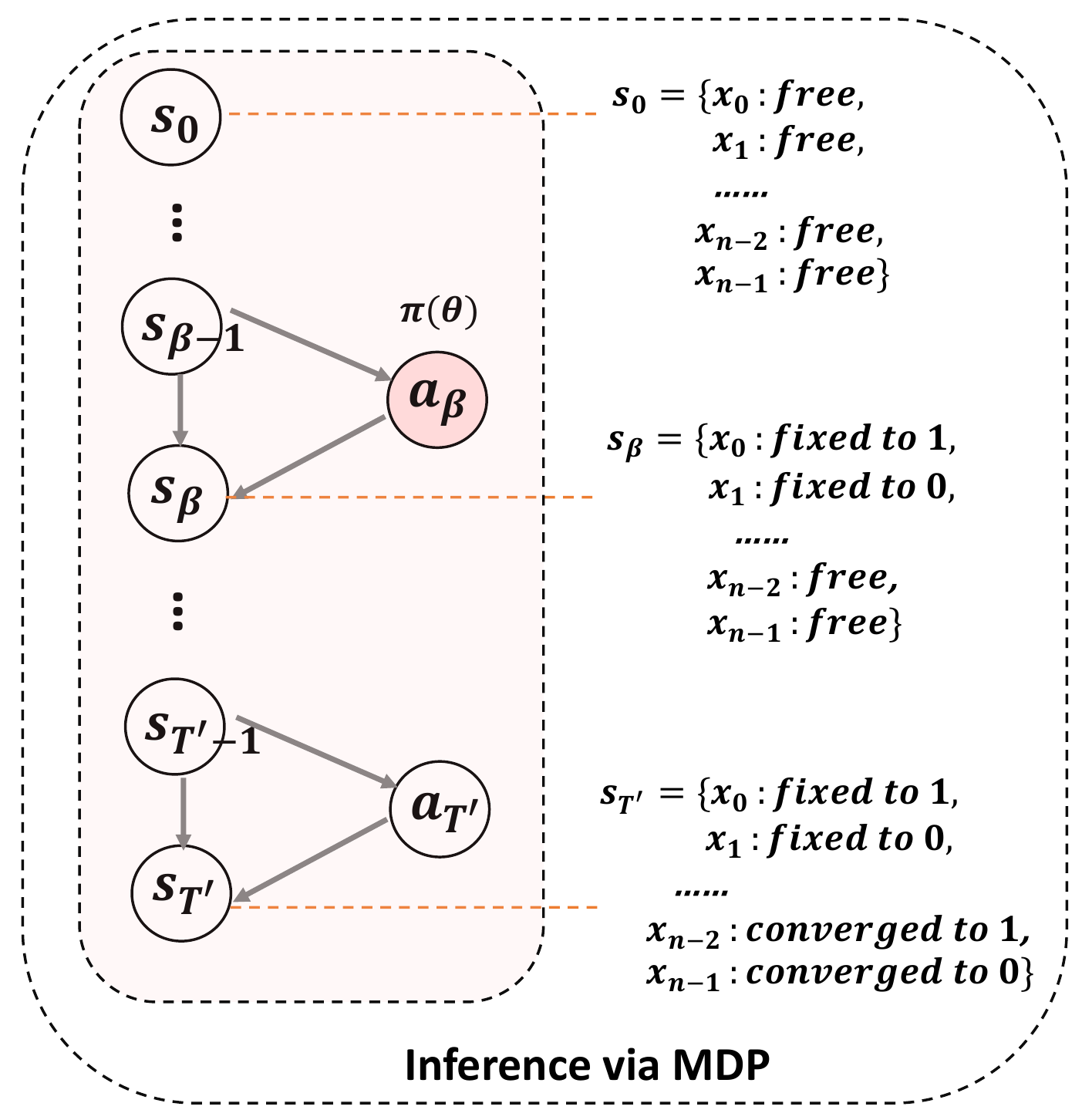}
  \caption{Solving an instance with Early Fixing via Markov decision process.}
  \label{fig_ef} %用于文内引用的标签
\end{figure}

\vspace{-1em}
\subsection{Markov decision process formulations}\label{mdp}
The sequential decisions made in each early-fixing process actually construct a Markov decision process\cite{howard1960dynamic}, as shown in Fig. \ref{fig_ef}. Considering the approximate method $\mathcal{M}$ to be accelerated as the environment, we define the states, actions, and policy for early fixing as following:

\begin{itemize}
\item \textbf{\textit{States}}: state $\bm{s}_t$ ($t \in \{0,...,\textit{T'}\}$) denotes the state at the iteration $t$, which is a state set of all variables. The state of each variable $\bm{s}_{t,i}$ could be anyone out of the five possible states: being fixed/converged to 1, being fixed/converged to 0, or staying free. We denote $\bm{s}_t = \{\bm{s}_{t,i}\}_{i=0}^{n{-}1}$. The initial state $\bm{s}_0$ exhibits that all the variables are free, while the terminate state $\bm{s}_{T'}$ represents that all the variables are fixed/converged, no matter to 0 or 1. 

\item \textbf{\textit{Actions}}: the action $\bm{a}_t$ ($t \in \{1,...,\textit{T'}\}$) denotes the transitions from one state to the next state. Since our early fixing process is carried out once every $\beta$ iterations, then the action is also conducted once every $\beta$ iterations. The terminate iteration could trigger the actions because the variables converge. $\bm{a}_t$ is also a set of actions of each variable: $\bm{a}_t = \{{a}_{t,i}\}_{i=0}^{n{-}1}$. 

\item \textbf{\textit{Policy}}: the policy $\pi(\bm{a}_t | \bm{s}_t)$, is given by the policy network as described in subsection \ref{learn2fix}, which inputs the past $\beta$ iteration values of all the currently free variables, and outputs the probabilities of these free variables whether to fix them or not. We use $\bm{p}$ to denote the output: $\bm{p} = \pi(\bm{a}_t | \bm{s}_t)$ where $p_i \in [0,1], i \in \{1,...,u\}$, $u$ is the number of free variables. We set a fixing threshold $\delta \in [0.5,1]$. If $p_i > \delta$, early fix this variable to 1; If $p_i < (1-\delta)$, early fix this variable to 0; Otherwise, do not conduct early fixing and this variable stays free. 

\end{itemize}

\begin{figure*}[!t] %H为当前位置，!htb为忽略美学标准，htbp为浮动图形
  \centering %图片居中
  \includegraphics[width=0.90\textwidth]{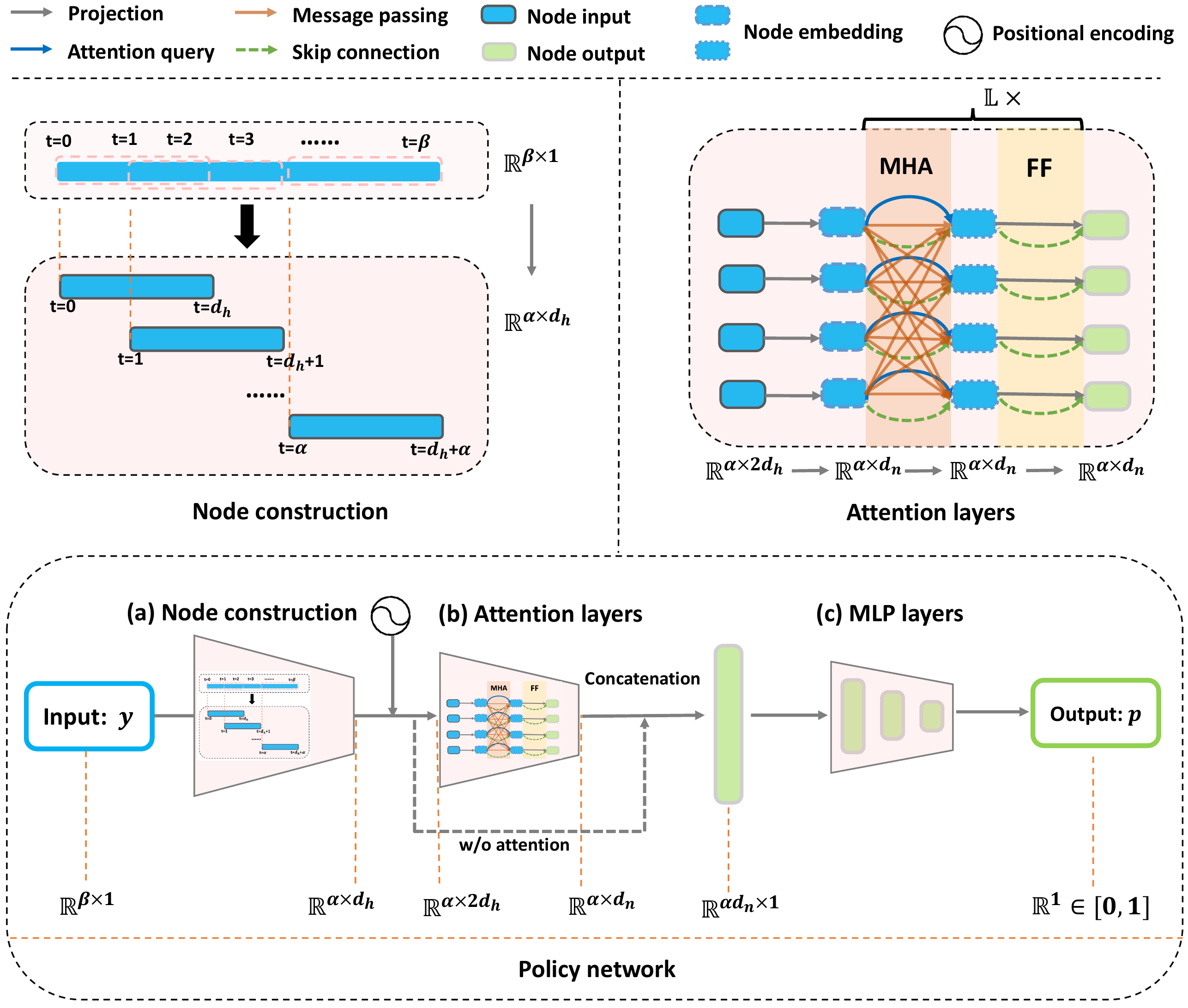}
  \caption{Policy network: the input is with dimension $\beta$. \textbf{(a)} After node construction, there are $\alpha$ nodes, each with dimension $d_h$. With the positional encoding, the dimension turns to $2d_h$. \textbf{(b)} Then go through $\mathbb{L}$ attention layers, including a multi-headed attention (MHA) sublayer and a feed-forward (FF) sublayer. \textbf{(c)} The nodes are concatencated as one vector, and then fed into the multi-layer perceptron (MLP) layers. Finally, the output is a probability $p \in [0,1]$, determining whether to fix the input variable or not, according to its past $\beta$ iterations.}
  \label{fig3} %用于文内引用的标签
\end{figure*}

\noindent According to the policy $\pi(\bm{a}_t | \bm{s}_t)$, we may update the variable states, converting some variables from free to fixed. The iteration will terminate when the stopping criterion reaches convergence, or all the variables have been fixed. As a Markov decision process, this early fixing process is episodic, where each episode amounts to solving the IP problem. An ideal approach to find an early-fixing policy is reinforcement learning, here we adopt an imitation learning scheme to train the network. Instead of training the network from scratch, we learn the features of early iterations and directly map to the converged solutions in order to achieve the purpose of acceleration.

% \vspace{-1em}
\subsection{Policy network with attention}
\label{learn2fix}

The policy network provides with the probability for determining whether to fix the input variable or not at current state. The input of the policy network is the past $\beta$ iterative values of the given variable, and the output is a probability $p \in [0,1]$. Taking one variable ${y}$ as an example, we firstly collect the past $\beta$ continuously iterative values of the variable, which are sequential according to the time series. At this point, the dimension is $\beta\times1$. Inspired by the sequential model \cite{chien2021hierarchical}, we use the sliding window to convert the $\bm{{y}}\in \mathbb{R}^{\beta\times1}$ to $\bm{{z}}\in \mathbb{R}^{\alpha\times{d_h}}$, where $\bm{{z}}$ is the iteration embedding of $\bm{y}$, $\alpha$ denotes the node number, and $d_h$ is the node dimension.

\noindent\textbf{Positional encoding}  
There is no recurrence and no convolution in the attention models, so we need positional encoding. Before feeding $\bm{z}$ into the attention layers, we inject some information about the relative position of the tokens in the sequence of $\bm{z}$, so as to make the most use of the order of the sequence. We follow the setup in Vaswani et al. \cite{vaswani2017attention} and add the positional encodings to the iteration embeddings. Let $k$ be the node number, $d_h$ be the embedding dimension. Then the positional encodings have the same dimensions $d_h$ as the embeddings. We use sine and cosine functions of different frequencies:

\vspace{-0.5em}
\begin{align}
  \label{eq4}
  \begin{cases}
  PE_{(k,2j)} &= sin{\left(k/10000^{2j/d_h}\right)}  \\
  PE_{(k,2j{+}1)} &= cos{\left(k/10000^{2j/d_h}\right)},
  \end{cases}
 \end{align}

\noindent where \textit{k} is the position, \textit{k}${\in} \{1,...,\alpha\}$ and \textit{j} is the dimension, \textit{2j} $\leq d_h$. The wavelengths form a geometric progression from $2\pi$ to $10000{\times} 2\pi$. After adding the positional encoding, the iteration embedding turns from $\bm{{z}}\in \mathbb{R}^{\alpha\times{d_h}}$ to $\bm{{\hat{z}}}\in \mathbb{R}^{\alpha\times{2d_h}}$.

\noindent\textbf{Attention layers} \ Then we apply the encoder part of Transformer-alike attention architecture \cite{kool2018attention} to our network to extract better iteration embeddings. First of all, to make it consistent with the dimensions, through a learned linear projection, one node input $\bm{\hat{z}}_k$ is projected to one node embedding $\bm{h}_k$, where $\bm{\hat{z}}_k \in \mathbb{R}^{2d_h}$ and $\bm{h}_k \in \mathbb{R}^{d_n}$, and the dimension $d_n$=128. Then all the node embeddings go through $\mathbb{L}$ attention layers. Each layer consists of two sublayers: a multi-head attention (MHA) layer that executes message passing between the nodes, and a node-wise fully connected feed-forward (FF) layer. Each sublayer adds a skip-connection \cite{he2016deep} and a batch normalization (BN) \cite{ioffe2015batch}:

\begin{figure}[!t] %H为当前位置，!htb为忽略美学标准，htbp为浮动图形
  \centering %图片居中
  \includegraphics[width=0.43\textwidth]{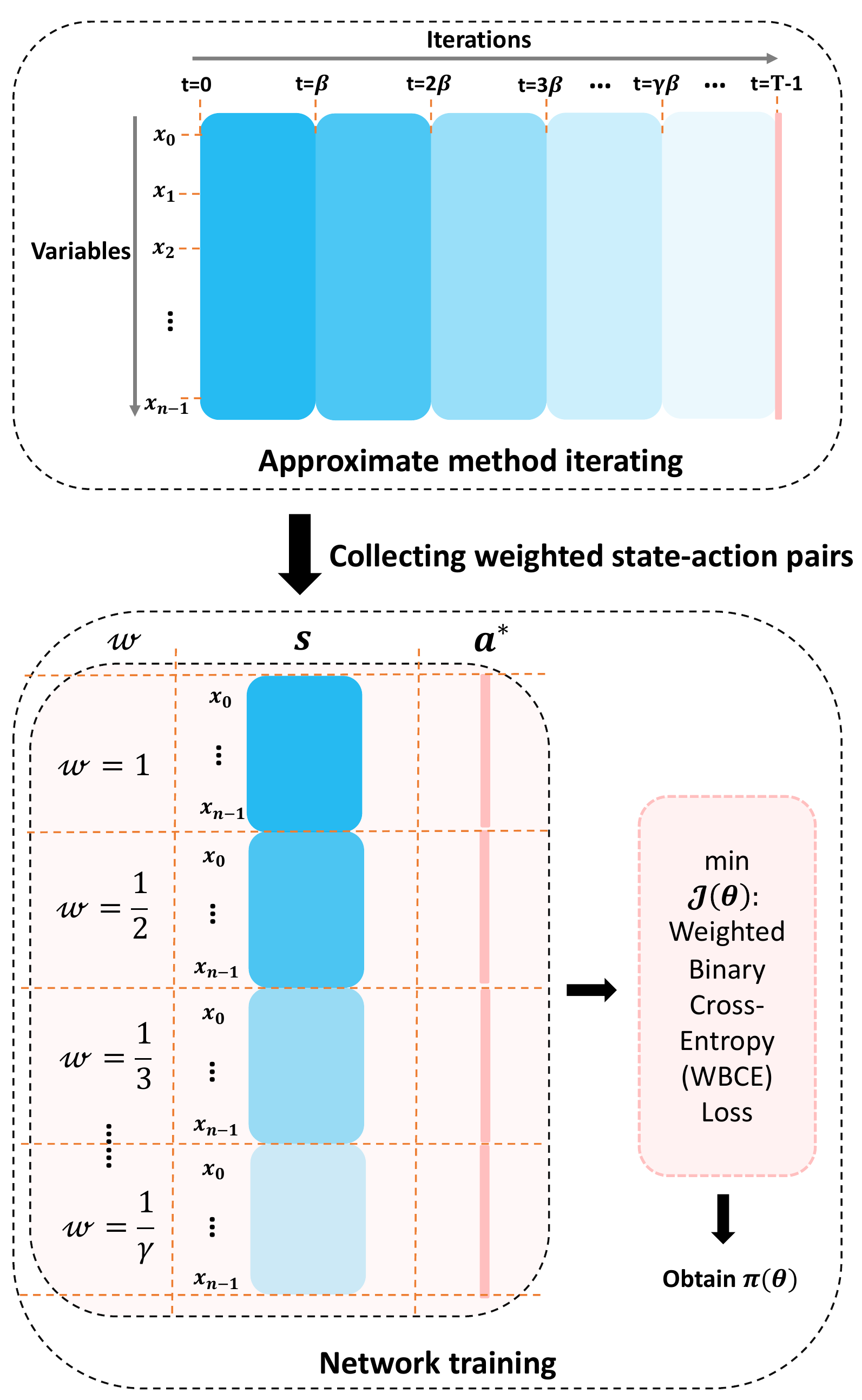}
  \caption{Network training with weighted binary cross-entropy (WBCE) loss.}
  \label{fig_train} %用于文内引用的标签
\end{figure}

\begin{align}
  \label{eq4}
  \begin{cases}
%   \mathop{\arg\max}_{\textbf{x}} \ \textbf{c}^T\textbf{x}, \ \  s.t. \ \ 
  \bm{\hat{h}}_k \ \ = {\rm{BN}}^{\ell} \left( \bm{\hat{h}}_k^{(\ell-1)}+{\rm{MHA}}_k^{\ell}\left(\bm{\hat{h}}_1^{(\ell-1)},...,\bm{\hat{h}}_\alpha^{(\ell-1)}\right) \right) \\
  \bm{h}_k^{(\ell)} = {\rm{BN}}^{\ell} \left( \bm{\hat{h}}_k+{\rm{FF}}^{\ell}(\bm{\hat{h}}_k) \right). 
  \end{cases}
 \end{align}

\noindent Any two layers do not share their parameters. Layer index $\ell \in \{1,..\mathbb{L}\}$, node index $k \in \{1,...,\alpha\}$. The MHA sublayer uses $H=8$ heads, each with dimension $\frac{d_n}{H}=16$. Moreover, the fully connected FF sublayer, which is applied to each node embedding separately and identically, consists of two linear transformations with a ReLU activation function in between: it first maps the node embedding from dimension $d_n$ to hidden dimension $d_{n'}=512$, then transforms from $d_{n'}$ back to $d_n$. After $\mathbb{L}$ attentions layers, we get $\alpha$ node embeddings, each with dimension $d_n$. At this point, we do the concatenation and get embedded variable $\bm{\overline{z}}$ with dimension $(\alpha{\cdot}{d_n}) \times 1$.

\noindent\textbf{MLP layers} \ After the attention layers, the variable $\bm{y}$ obtains a new embedding $\bm{\overline{z}} \in \mathbb{R}^{(\alpha{\cdot}{d_n}) {\times} 1}$, which will be fed into another multi-layer perceptron (MLP). At this stage, we utilize three fully connected FF sublayers along with decreasing dimensions ($256-128-16$) with the ReLU activation functions between the hidden layers. Finally, in the last layer through a sigmoid function, we obtain a probability $p_y \in [0,1]$, determining whether to fix the input variable ${y}$ or not. 

If the probability is greater than the fixing threshold $\delta$, then the action of fixing this variable to 1 will be conducted. we can interpret this fixing threshold as a symmetric fixing confidence, \ie, if the probability is less than $1-\delta$, then the action of fixing this variable to 0 will be conducted. Otherwise, no fixing action will be conducted and this variable will be further updated. The early fixing process is given in the Algorithm \ref{alg1}.

\subsection{Training: imitation learning}\label{training}
% Our motivation for accelerating the approximate iterative methods for solving IP is simple: using the past $b$ iterations of variables to predict their finally converged solutions, and early fixing them so as to accelerate the optimization process.  \cite{gasse2019exact}
We train the policy network $\pi(\theta)$ as shown in Fig. \ref{fig_train}. Specifically, our policy network is trained by behavioral cloning \cite{torabi2018behavioral} as a method of expert-driven imitation learning, and here we use the approximate method to be accelerated $\mathcal{M}$ itself as the expert rule. Subsection \ref{mdp} has explained the meaning for states and actions. Then mathematically, for those free variables, we assign $\bm{s}$ as the past $\beta$ iterative values, and assign $\bm{a}^{*}$ as the ultimately converged discrete solutions of the approximate method $\mathcal{M}$, \ie, the expert solutions. We first run the expert on a collection of $N$ training instances, and pick the dataset of expert state-action pairs $\mathcal{D} {=} \{(\bm{s}_{e,r,i},{a}^{*}_{e,r,i} )|_{i=0}^{n{-}1}|_{r{=}0}^{\gamma{-}1}|_{e=0}^{N{-}1}\}$. And the policy is learned by minimizing the weighted binary cross-entropy (WBCE) loss:

\begin{align}
  \label{eq15}
  \mathcal{J}(\theta) = - \frac{1}{N{\cdot}\gamma{\cdot}n} \sum_{e=0}^{N{-}1} \sum_{r=0}^{\gamma{-}1} \sum_{i=0}^{n{-}1} w_{e,r,i}q_{e,r,i},
\end{align}

\begin{align}
  \label{eq15}
  q_{e,r,i} = {a}^{*} \log\pi_\theta(a|\bm{s}) + (1{-}{a}^{*}) \log(1{-}\pi_\theta(a|\bm{s})),
\end{align}

\noindent where for $a^*, a, \bm{s}$, we hide the subscripts $_{e,r,i}$ for readability. $w_{e,r,i}$ is the weight, $w_{e,r,i}=\frac{1}{r+1}$. Mathematically, $\bm{s}_{e,r,i} = \bm{x}^{(r,e)}_{(r{-}1)\beta:r\beta}$. ${a}^{*}_{e,r,i} = x^{(r,e)}_{T{-}1}$. We call one problem shown as in Formulation (\ref{eq1}), as one instance. $N$ is the number of training instances.  

% \vspace{-1em}
\subsection{Inference with early fixing} 
\label{updating}

The general inference stage of early fixing framework is given in the Algorithm \ref{alg1}, and we also present the process as a MDP in Fig. \ref{fig_ef}. Our early fixing framework takes each variable independently, and decisions on whether to early fix are once every block of iterations. In each block of iterations, given the iterative values of the variables within the past $\beta$ iterations, the policy network will evaluate the posterior probability of each variable concerning all discrete candidate states (0 or 1). If the posterior probability with respect to one state exceeds a threshold, namely the fixing threshold $\delta$, then the action of fixing this variable to that discrete state will be conducted, and this variable will not be updated in later iterations; otherwise, no fixing action will be conducted and this variable will be further updated. 

When a certain number of variables are fixed in previous iterations, then how to update the problem into a smaller-sized one will be discussed in this subsection. Our early fixing framework is available for both linear programming and quadratic programming, no matter constrained or unconstrained. We will give the mathematical assumptions and propositions based on a constrained quadratic programming problem: 
\begin{align}
    \label{eq5}
  \mathop{\arg\max}_{\textbf{x}} \ \textbf{x}^\top\textbf{A}\textbf{x} {+} \textbf{b}^\top\textbf{x}, \ \  s.t. \ \ \textbf{Cx} \bigotimes \textbf{d}, \ \ \textbf{x}{\in}\{0,1\}^n.
\end{align}

\noindent\textbf{Notations.} We denote the matrices and vectors in formulation (\ref{eq5}) as: 
\textbf{A} = $\begin{bmatrix} \bm{A}_1 & \bm{A}_2 \\ \bm{A}_3 & \bm{A}_4 \end{bmatrix}, \textbf{C} = \begin{bmatrix} \bm{C}_1 & \bm{C}_2 \end{bmatrix}$, \textbf{b} = $\begin{bmatrix} \bm{b}_1 \\ \bm{b}_2 \end{bmatrix}$, \textbf{x} = $\begin{bmatrix} \textbf{x}_1 \\ \textbf{x}_2 \end{bmatrix}$, where vectors $\textbf{x}_1$ refers to the set of free variables and $\textbf{x}_2$ refers to the set of fixed variables, and the same around for other vectors and matrices. Let $u,v$ be the number of free and fixed variables, then $\bm{b}_1, \textbf{x}_1 {\in} \mathbb{R}^u, \bm{b}_2, \textbf{x}_2 {\in} \mathbb{R}^v$, $\bm{A}_1 {\in} \mathbb{R}^{u\times u}$, $\bm{A}_2 {\in} \mathbb{R}^{u\times v}$, $\bm{A}_3 {\in} \mathbb{R}^{v\times u}$, $\bm{A}_4 {\in} \mathbb{R}^{v\times v}$, $\bm{C}_1 {\in} \mathbb{R}^{m\times u}$, $\bm{C}_2 {\in} \mathbb{R}^{m\times v}$, $\bigotimes$ denotes any relational symbol such as $<, >, =, \geq$ or $\leq$. 

\begin{algorithm}[!t]
    % \label{alg1}
    \begin{algorithmic}[1]
    % \label{alg1}
      \caption{Inference: Early Fixing Framework}\label{alg1}
      \REQUIRE accelerated approximate method $\mathcal{M}$, instance $\mathcal{I}$, policy network $\pi_\theta$, total variable number $n$, block size $\beta$, fixing threshold $\delta \in [0.5,1]$  %free variable number $u$, fixed variable number $v$, 
      \ENSURE $\bm{x}^*$
      \STATE $u\leftarrow n, v\leftarrow 0, t\leftarrow \beta$
    %   \STATE Initialize $\bm{x} \leftarrow \bm{0}.$ 
      \REPEAT 
      \IF{$t \% \beta == 0$}
      \STATE $\bm{x}_{t{-}\beta:t} \leftarrow \mathcal{M}(\mathcal{I})$  
      \STATE $\bm{s}_t \leftarrow \bm{x}_{t{-}\beta:t}$  
      \STATE $\bm{p} \leftarrow \pi_\theta(\bm{a}_t|\bm{s}_t)$
      \FOR{$i=1$ to $u$}
      \IF{$p_i > \delta$}
      \STATE Early fix variable $x_i$ to 1, $v\leftarrow v{+}1$
      \ENDIF 
      \IF{$p_i < 1 {-} \delta$}
      \STATE Early fix variable $x_i$ to 0, $v\leftarrow v{+}1$
      \ENDIF 
      \ENDFOR 
      \STATE Update the instance $\mathcal{I}$ as Subsection \ref{updating}. 
      \STATE $u \leftarrow u{-}v$, $v\leftarrow0, t\leftarrow t{+}\beta$
      \ENDIF 
      \UNTIL{(\textit{converged} or $u \leq 0$)}
      \STATE Record all the fixed/converged discrete solutions $\bm{x}^*$
      \STATE \textbf{return} $\bm{x}^*$.
      
    %   \FOR{$r=1$ to $T$}
    %   \STATE $x_{t} \leftarrow M(\cdot)$
    %   \IF{stopping criteria satisfied OR $u==0$}
    %   \STATE Break
    %   \ENDIF 
    %   \STATE Append $x_t$ to the iteration list $xIters$
    %   \IF{$t \ \% \ b == 0$}
    %   \STATE $p_{(1:u)} \leftarrow Q(xIters)$
    %   \FOR{$i=1$ to $u$}
    %   \STATE Early fix $x_i\leftarrow$  
    %   \begin{cases}
    %      1, \ if: p_i \geq C\\
    %      0, \ if: p_i \leq (1-C)\\
    %   \end{cases}, $f{\leftarrow}f+1$ 
    %   \STATE No early fix, otherwise
    %   \ENDFOR 
    %   \STATE $u{\leftarrow}u{-}f$, $f{\leftarrow}0$
    %   \STATE Update the parameters in $M$, and reset $xIters{\leftarrow}[ \ ]$
    %   \ENDIF
    %   \ENDFOR
    %   \STATE obtain $x^*$ by the set of ultimately fixed and unfixed variables
    %   \STATE \textbf{return} $\bm{x}$
    
    \end{algorithmic}
\end{algorithm}

% \begin{algorithm}[!t]
%     % \label{alg1}
%     \begin{algorithmic}[1]
%     % \label{alg1}
%       \caption{Inference: Early Fixing Framework}\label{alg1}
%       \REQUIRE the approximate method to be accelerated $\mathcal{M}$, policy network $\pi$, total variable number $n$, free variable number $u$, fixed variable number $v$, block size $\beta$, fixing confidence $C \in [0.5,1)$ 
%       \ENSURE $x^{*}$
%       \STATE Initialize the iteration list to empty: $xIters{\leftarrow}[ \ ]$
%       \STATE $u\leftarrow n, v\leftarrow 0$
%       \FOR{$t=1$ to $T$}
%       \STATE $x_{t} \leftarrow M(x_{t-1})$
%       \IF{stopping criteria satisfied OR $u==0$}
%       \STATE Break
%       \ENDIF 
%       \STATE Append $x_t$ to the iteration list $xIters$
%       \IF{$t \ \% \ b == 0$}
%       \STATE $p_{(1:u)} \leftarrow Q(xIters)$
%       \FOR{$i=1$ to $u$}
%       \STATE Early fix $x_i\leftarrow$  
%       \begin{cases}
%          1, \ if: p_i \geq C\\
%          0, \ if: p_i \leq (1-C)\\
%       \end{cases}, $f{\leftarrow}f+1$ 
%       \STATE No early fix, otherwise
%       \ENDFOR 
%       \STATE $u{\leftarrow}u{-}f$, $f{\leftarrow}0$
%       \STATE Update the parameters in $M$, and reset $xIters{\leftarrow}[ \ ]$
%       \ENDIF
    
%       \ENDFOR
%       \STATE obtain $x^*$ by the set of ultimately fixed and unfixed variables
%       \STATE \textbf{return} $x^*$
    
%     \end{algorithmic}
% \end{algorithm}  

\noindent\textbf{Proposition 1.}  Problem reformulation: when doing the early fixing once every $\beta$ iterations, let $r$ be the rounds for conducting early fixing, then we can propose that:
\begin{itemize}
    \item $\textbf{x}^{(r+1)}$ = $\textbf{x}^{(r)}_1$
    \item $\bm{A}^{(r+1)}$ = $\bm{A}^{(r)}_1$
    \item $\bm{b}^{(r+1)}$ = $(\bm{A}_2^{(r)}+\bm{{A}_3}^{\top(r)})\textbf{x}^{(r)}_2 + \bm{b}^{(r)}_1$
    \item $\bm{C}^{(r+1)}$ = $\bm{C}^{(r)}_1$
    \item $\bm{d}^{(r+1)}$ = $\bm{d}^{(r)}-\bm{C}_2^{(r)}\textbf{x}_2^{(r)}$
\end{itemize}

\noindent\textbf{Proof 1.} We eliminate the superscript $^{(r)}$ for readability.  
\begin{itemize}
\item[(i)] For the objective function:  
\begin{align}
  & \qquad \textbf{x}^\top\textbf{A}\textbf{x} + \textbf{b}^\top\textbf{x} \\
  &= \begin{bmatrix} \textbf{x}_1^\top & \textbf{x}_2\top \end{bmatrix}  \begin{bmatrix} \bm{A}_1 & \bm{A}_2 \\ \bm{A}_3 & \bm{A}_4\end{bmatrix} \begin{bmatrix} \textbf{x}_1 \\ \textbf{x}_2 \end{bmatrix} + \begin{bmatrix} \bm{b}_1^\top & \bm{b}_2^\top \end{bmatrix} \begin{bmatrix} \textbf{x}_1 \\ \textbf{x}_2 \end{bmatrix} \\
  &= \textbf{x}_1^\top \bm{A}_1 \textbf{x}_1 + \textbf{x}_2^\top \bm{A}_3 \textbf{x}_1 + \textbf{x}_1^\top \bm{A}_2 \textbf{x}_2 \\ & \qquad\qquad\qquad\qquad\quad + \textbf{x}_2^\top \bm{A}_4 \textbf{x}_2 + \bm{b}_1^\top \textbf{x}_1 + \bm{b}_2^\top \textbf{x}_2 \nonumber \\
  &= \textbf{x}_1^\top \bm{A}_1 \textbf{x}_1 + ((\bm{A}_2+\bm{A}_3^\top)\textbf{x}_2 + \bm{b}_1)^\top\textbf{x}_1 \\ & \qquad\qquad\qquad\qquad\qquad + (\textbf{x}_2^\top\bm{A}_4\textbf{x}_2+\bm{b}_2^\top\textbf{x}_2),\nonumber 
\end{align}

thus: $\textbf{x}^{(r+1)}$ = $\textbf{x}^{(r)}_1$, $\bm{A}^{(r+1)}$ = $\bm{A}^{(r)}_1$, $\bm{b}^{(r+1)}$ = $(\bm{A}_2^{(r)}+\bm{A}_3^{\top(r)})\textbf{x}^{(r)}_2 + \bm{b}^{(r)}_1$. Since previously fixed variables can be seen as the constants in the following iterations, then ($\textbf{x}_2^\top\bm{A}_4\textbf{x}_2+\bm{b}_2^\top\textbf{x}_2$) is constant. 

\item[(ii)] For the constraints: 
\begin{align}
  \textbf{Cx} & \bigotimes \textbf{d} \\
  \begin{bmatrix} \bm{C}_1 & \bm{C}_2 \end{bmatrix} \begin{bmatrix} \textbf{x}_1 \\ \textbf{x}_2 \end{bmatrix} &\bigotimes \bm{d} \\
  \bm{C}_1\textbf{x}_1 &\bigotimes \bm{d} - \bm{C}_2 \textbf{x}_2
\end{align}

thus: $\bm{C}^{(r+1)}$ = $\bm{C}^{(r)}_1$, $\bm{d}^{(r+1)}$ = $\bm{d}^{(r)}-\bm{C}_2^{(r)}\textbf{x}_2^{(r)}$.
\end{itemize}
\noindent  Proof ends. 

\noindent\textbf{Remark 1.} As regard to the updating for $\bm{b}$, if $\bm{A}$ is a symmetric matrix, then $\bm{A}_2 = \bm{A}_3^\top$. At that point, the updating can be simplified as: $\bm{b}^{(r+1)}$ = $2\bm{A}_2^{(r)}\textbf{x}^{(r)}_2 + \bm{b}^{(r)}_1$.

% \subsection{Hand-designed function to early fix}
% The variables will be fixed, if the variable iterations values $xIters$ stay almost unchanged in 3 consistent iterations. The mathematical formulation is given as: 
% \begin{align}
%   \label{eq5}
%   ( \sum_{k=k+1}^{k+3} (x_{k+1} - x_{k}) \leqslant fixThreshold \ )=3,  \ where \ k \in [i \times W,(i+1) \times W - 3] 
% \end{align}

\section{Constrained Linear Programming}
\label{linear}

\begin{table*}[!t]
\label{tab1}
    \centering
\caption{Performance evaluations for constrained linear programming on generated regular-sized Dataset \uppercase\expandafter{\romannumeral1}. The time limit is set to 1 hour.} %, training on 12800 instances and test on 1000 instances. The method RL+LSTM is by \cite{pan2020solving}
\label{tab1}
\resizebox{16.8cm}{!}{
\begin{tabular}{lcccccccccccc}
    \toprule[1.5pt]
     Size $\rightarrow$ & \multicolumn{3}{c}{$n=500$} & \multicolumn{3}{c}{$n=1000$} & \multicolumn{3}{c}{$n=1500$}  & \multicolumn{3}{c}{$n=4000$} \\
     \cmidrule(lr){2-4}\cmidrule(lr){5-7}\cmidrule(lr){8-10}\cmidrule(lr){11-13}
     Model & Obj.$\uparrow$ & Obj. Gap$\downarrow$ & Time$\downarrow$ & Obj.$\uparrow$ & Obj. Gap$\downarrow$ & Time$\downarrow$ & Obj.$\uparrow$ & Obj. Gap$\downarrow$ & Time$\downarrow$ & Obj.$\uparrow$ & Obj. Gap$\downarrow$ & Time$\downarrow$\\
    \cmidrule(lr){1-4}\cmidrule(lr){5-7}\cmidrule(lr){8-10}\cmidrule(lr){11-13}
    RPB \cite{achterberg2009hybrid}&  7464.8 & N/A & 2.79s & 14888& N/A	& 23.17s	& 21231 & N/A & 169.16s & 59772 & N/A & 3600s \\
	FiLM \cite{gupta2020hybrid}&  7464.8 & N/A & 2.08s & 14888& N/A	& 19.78s & 21231 & N/A & 185.79s & 58809 & N/A & 3600s   \\
    GCNN \cite{gasse2019exact}&  \textbf{7464.8} & N/A & \textbf{1.78s} & \textbf{14888}& N/A	& \textbf{15.22s} & \textbf{21231} & N/A & \textbf{144.74s} & \textbf{60105} & N/A & 3600s   \\
    
    \cmidrule(lr){1-4}\cmidrule(lr){5-7}\cmidrule(lr){8-10}\cmidrule(lr){11-13}
    $\ell_p$-box ADMM \cite{wu2019lp} &  \textbf{6953.7} & N/A & 1.17 & \textbf{14430} & N/A & 2.54s & \textbf{20930} & N/A & 4.04s & \textbf{56188} & N/A & 11.15s    \\
    % $\ell_p$box+ES &  5638.91 & 24.46\% & 0.63 & 11240.05 & 24.50\% & 1.32 & 16698.81 & 24.88\% & 2.08 & 44712.11 & 25.61\% & 5.78 \\
    % $\ell_p$-box ADMM + HEF & 6204.2 & 10.77\% & 0.24s & 12590 & 12.75\% & 0.45s & 18125 & 13.40\% & 0.44s & 48324 & 13.99\% & 2.26s   \\
    $\ell_p$-box ADMM + LEF(w/o Att.) & {6804.9} & \textbf{0.69\%} & \textbf{0.18s} & 13768 & 4.10\% & \textbf{0.38s} & 20415 & 2.40\% & \textbf{0.43s} & 53195 & 5.28\% & \textbf{1.39s}    \\
	$\ell_p$-box ADMM + LEF(with Att.) & 6883.5 & 1.96\% & 0.27s & 13803 & \textbf{3.96\%} & {0.44s} & 20500 & \textbf{1.85\%} & 0.56s & 54020 & \textbf{3.81\%} & 2.47s \\

    \bottomrule[1.5pt]
\end{tabular}}
\end{table*}

\begin{table*}
    \centering
\caption{Performance evaluations for constrained linear programming on generated large-sized Dataset \uppercase\expandafter{\romannumeral2}. A negative objective gap means a better objective.} %, training on 12800 instances and test on 1000 instances. The method RL+LSTM is by \cite{pan2020solving}
\label{tab2}
\resizebox{17.9cm}{!}{
\begin{tabular}{lcccccccccccc}
    \toprule[1.5pt]
     Size $\rightarrow$ & \multicolumn{6}{c}{$n=1e4$} & \multicolumn{6}{c}{$n=5e4$} \\ 
     \cmidrule(lr){2-7}\cmidrule(lr){8-13} 
    Model & Obj.$\uparrow$ & Obj. Gap$\downarrow$ & Time$\downarrow$ & Speedup$\uparrow$ & \#Sol. Diff.$\downarrow$ & Accuracy$\uparrow$ & Obj.$\uparrow$ & Obj. Gap$\downarrow$ & Time$\downarrow$ & Speedup$\uparrow$ & \#Sol. Diff.$\downarrow$ & Accuracy$\uparrow$ \\
    \cmidrule(lr){1-7}\cmidrule(lr){8-13}  
    $\ell_p$-box ADMM \cite{wu2019lp}& 9665.8 & N/A & 11.4s & N/A & N/A & N/A & 48372 & N/A & 111.5s & N/A & N/A & N/A     \\
    % $\ell_p$-box ADMM + HEF & {9565.3} & 1.03\% & 3.7s & 3.1\times  & 22.7 & 99.7730\% &  {47928} & 0.92\% & 38.5s & 3.9\times & 131.8 & 99.7364\%   \\
    $\ell_p$-box ADMM + LEF(w/o Att.) & 9691.4 & -0.25\% & $\textbf{0.9s}$ &$\textbf{12.6}\times$ & 10.9 & 99.8910\% & 48445 & -0.15\% & $\textbf{5.4s}$ & $\textbf{20.6}\times$ & 66.0 & 99.8680\% \\
    $\ell_p$-box ADMM + LEF(with Att.) & $\textbf{9691.6}$ & $\textbf{-0.26\%}$ & $\textbf{0.9s}$ & $\textbf{12.6}\times$ & $\textbf{9.1}$ & $\textbf{99.9090\%}$ & $\textbf{48465}$ & $\textbf{-0.19\%}$ & 5.7s & 19.6$\times$ & $\textbf{55.0}$ & $\textbf{99.8900\%}$ \\
    \cmidrule(lr){1-7}\cmidrule(lr){8-13}

    Size $\rightarrow$ & \multicolumn{6}{c}{$n=1e5$} & \multicolumn{6}{c}{$n=2e5$} \\ 
    \cmidrule(lr){2-7}\cmidrule(lr){8-13} 
    Model & Obj.$\uparrow$ & Obj. Gap$\downarrow$ & Time$\downarrow$ & Speedup$\uparrow$ & \#Sol. Diff.$\downarrow$ & Accuracy$\uparrow$ & Obj.$\uparrow$ & Obj. Gap$\downarrow$ & Time$\downarrow$ & Speedup$\uparrow$ & \#Sol. Diff.$\downarrow$ & Accuracy$\uparrow$ \\
    \cmidrule(lr){1-7}\cmidrule(lr){8-13}  
    $\ell_p$-box ADMM \cite{wu2019lp}&  97579 & N/A & 327.3s & N/A & N/A & N/A & 195445 & N/A & 990.0s & N/A & N/A & N/A \\
    % $\ell_p$-box ADMM + HEF & 96392 & 1.22\% & 56.4s & 5.8\times & 329.0 & 99.6710\% & 194431 & 0.51\% & 137.2s & 7.2\times & 545.0 & 99.7275\%  \\
    $\ell_p$-box ADMM + LEF(w/o Att.) & 97631 & -0.05\% & $\textbf{13.8s}$ & $\textbf{23.7}\times$ & 129.8 & 99.8702\% & 195710 & -0.14\% & $\textbf{39.4s}$ & $\textbf{25.1}\times$ & 247.1 & 99.8765\%  \\
    $\ell_p$-box ADMM + LEF(with Att.) & $\textbf{97682}$ & $\textbf{-0.11\%}$ & 14.4s & 22.7$\times$ & $\textbf{126.8}$ & $\textbf{99.8732\%}$ & $\textbf{195758}$ & $\textbf{-0.16\%}$ & 41.3s & 24.0$\times$ & $\textbf{220.2}$ & $\textbf{99.8899\%}$    \\
    \bottomrule[1.5pt]
\end{tabular}}
\end{table*}

\subsection{Setup}
\noindent \textbf{Accelerated method and datasets} \ In this section, we will accelerate the approximate method: $\ell_p$-box ADMM \cite{wu2019lp} for solving constrained linear integer programming. For the datasets, we choose the combinatorial auction problems \cite{gasse2019exact} with the following formulation:
\begin{align}
    \label{eq14}
  \mathop{\arg\max}_{\textbf{x}} \ \textbf{b}^\top\textbf{x}, \ \  s.t. \ \ \textbf{Cx} {\leqslant} \textbf{d}, \ \ \textbf{x}{\in}\{0,1\}^n, 
\end{align}
% \vspace{-0.5em}

\noindent where $\textbf{C} \in \mathbb{R}^{m\times n}$, $m,n$ refer to the number of constraints and variables. We follow the experimental setup of Gasse et al. \cite{gasse2019exact}, and generate two sets of instances in difference sizes. 
\begin{align}
    \label{eq15}
  \textbf{b,C,d} \leftarrow \mathbb{G}({\rm{bid}}{=}n,{\rm{item}}{=}\xi {\cdot} m).
\end{align}
We generate the instances, namely, $\textbf{b,C,d}$, according to the generator $\mathbb{G}$. The instance size is determined by two key parameters, bid and item. The bid number is equivalent to variable number $n$, while the item number is not equivalent but directly proportional to the constraint number $m$. $\xi$ is a constant. In order to ensure the feasibility and optimality of the instance, the constraint number $m$ could be different for different instances, given the same item number. We generate two sets of instances. For datasets \uppercase\expandafter{\romannumeral1}, there are regular-sized instances: (bid=500, item=100), (bid=1000, item=200), (bid=1500, item=300), (bid=4000, item=800). For datasets \uppercase\expandafter{\romannumeral2}, there are extra large scale instances: (bid=1e4, item=100), (bid=5e4, item=500), (bid=1e5, item=1000), (bid=2e5, item=2000).

\noindent\textbf{Training details} We train our model only on the smallest sized instances with $n$=500, and generalize to all other larger-sized instances. We train on $N=100$ instances, and do the inference on $N'=20$ instances for all the datasets. $\gamma, \beta, \delta, \mathbb{L}$ are set to 10, 100, 0.9, 2, respectively. We train for 10 epochs. The learning rate is set to $1e{-}4$. For our learning-based early fixing methods, training without attention for one epoch costs 77s, while training with attention costs 89s. We implement the functions of the accelerated methods in C++ and call the functions in Python via Cython interfaces. All the learning modules are implemented in Python.  

\noindent\textbf{Evaluations}  We evaluate all the different sizes, each with 20 instances. We set the time limit to 1 hour. For all datasets, we evaluate the objective, the objective gap and runtime, and we record the mean value of all instances. The objective gap is used to exhibit the gap between the $\ell_p$-box ADMM with learning-based early fixing (LEF) and the $\ell_p$-box ADMM without early fixing, given as: $\frac{obj_1 - obj_2}{obj_1}$, where $obj_1, obj_2$ refer to the objective obtained by $\ell_p$-box ADMM, $\ell_p$-box ADMM + LEF, respectively. A negative objective gap means achieving a better objective with early fixing. Specifically, for datasets \uppercase\expandafter{\romannumeral2}, we also evaluate the Speedup, the number of solution difference (\#Sol. Diff.) and the accuracy. The speedup is the time speedup, simply dividing the runtime of $\ell_p$-box ADMM by that of $\ell_p$-box ADMM + LEF. \#Sol. Diff is the number of solution difference where the variable solution by $\ell_p$-box ADMM is different from that by $\ell_p$-box ADMM + LEF, the maximum number should be the total variable number $n$. The accuracy is to evaluate the correct solution accuracy, given as: $\frac{n - n_d}{n} \times 100\%$, $n_d$ refers to (\#Sol. Diff.). In the hyperparameter study, we also evaluate the number of infeasible constraints. 
% When we early fix the variable, it is possible to fix it to the wrong solution, which may lead to the infeasible constraints. 

% \vspace{-0.1em}
\noindent\textbf{Baselines} For datasets \uppercase\expandafter{\romannumeral1}, we compare against three exact methods based on the branch-and-bound algorithm. Then for Datasets \uppercase\expandafter{\romannumeral2}, since the size is too large for them to obtain a solution, we only compare with the approximate methods. Reliability pseudocost branching (RPB) \cite{achterberg2009hybrid} is a variant of hybrid branching which is used by default in SCIP, and we choose SCIP $6.0.1$ \cite{achterberg2009scip} as the backend solver. Graph convolutional neural networks (GCNN) \cite{gasse2019exact} are applied to learning branch-and-bound variable selection policies. Feature-wise Linear Modulation (FiLM) \cite{gupta2020hybrid} \cite{perez2018film} layers are used to construct the neural network for learning to branch which is purely CPU-based, but shows competitive performances against GPU-based neural networks. 

% For the approximate method $\ell_p$-box ADMM \cite{wu2020map} to be accelerated, we compare our proposed learning based early fixing (LEF) against our hand-designed function based early fixing (HEF). We evaluate the efficiency of the attention layers. For the HEF, we set the rule as Proposition 2.

% \noindent\textbf{Proposition 2.}\label{pro2} Hand-designed early fixing (HEF) rule: if a variable remains unchanged ($|x_t{-}x_{t{-}1}|{\leq}{1e{-}4}$) in five consistent iterations, then early fix this variable to its binarized discrete state (zero or one). 

% early stopping (ES) at \textit{half} of the ultimately converged iterations. Besides, we compare with our hand-designed function (HF): we fix a variable if this variable remains almost unchanged ($\leq{1e{-}5}$) in three consistent iterations, as shown in equation (\ref{eq7}). We also compare our sequential network with attention model (AM) against the network with only the multi-layer perceptron (MLP), both with positional encodings. 

\subsection{Experimental results}
\label{analysis}

\begin{figure*}[!htb] %H为当前位置，!htb为忽略美学标准，htbp为浮动图形
    \centering %图片居中
    \includegraphics[width = 0.99\textwidth]{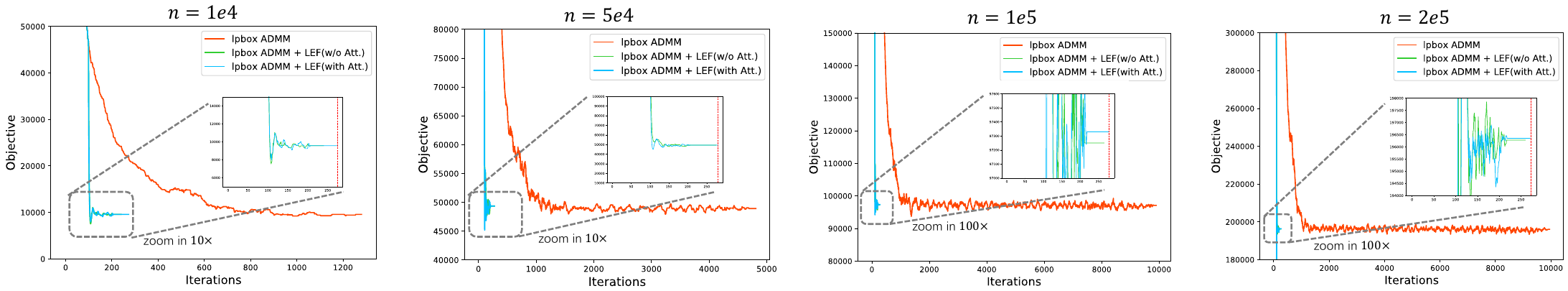}  
	\caption{Convergence on different methods for constrained linear programming: how the objective changes with respect to the iterations during the inference stages for four different sized instances. LEF method leads to much faster convergence, and LEF with attention layers achieves better objective.}
	\label{testing_lp} 
\end{figure*}

\noindent\textbf{Comparative study}
As shown in Table \ref{tab1} and Table \ref{tab2}, we firstly compare against three exact methods: RPB, FiLM and GCNN. GCNN generally outperforms RPB and FiLM with higher efficiency in runtime. The three exact methods have one obvious shortcoming in common: time-consuming, especially when the problem size increases, which is unacceptable in real life and large scale applications. Then, we compare our proposed LEF with the base method $\ell_p$-box ADMM. The results turn out that the LEF outperforms the base method regarding objective gaps, which reveals the effectiveness of our early fixing framework. In Fig. \ref{testing_lp}, we record how the objective changes with respect to the iterations during the inference stages for different sized instances. From the figure, we can see that our LEF method leads to much faster convergence. When zooming in $10\times$ in the left two figure, we obtain a general view about the LEF fluctuations. When zooming in $100\times$ in the right two figure, we can even clearly see that our LEF with attention layers achieves better objective.

\begin{figure*}[!htb] %H为当前位置，!htb为忽略美学标准，htbp为浮动图形
    \centering %图片居中
    \includegraphics[width = 0.99\textwidth]{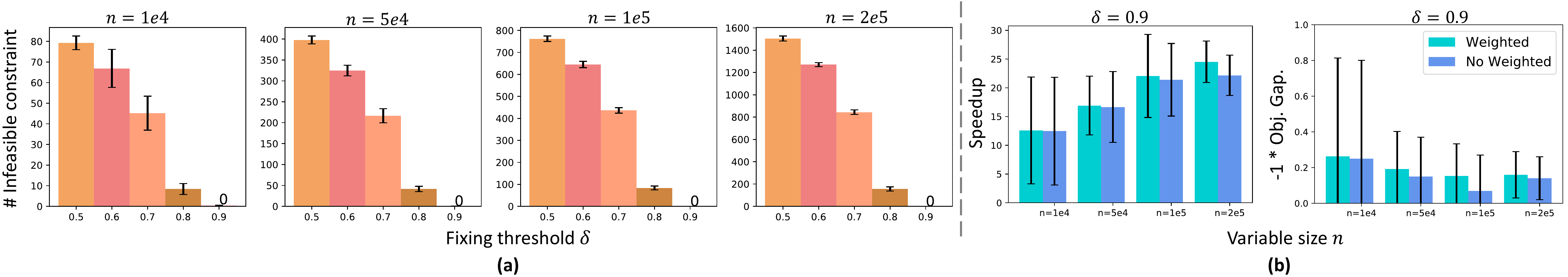} 
	\caption{Hyperparameter study on fixing threshold $\delta$ and ablation study on weighted loss: \textbf{(a)} How the number of infeasible constraints go with the fixing threshold. Different color refers to different fixing threshold. \textbf{(b)} How the objective gaps and runtime speedup of different sizes change when training with weighted loss or no weighted loss. No weighted loss means that all weights are equal to 1.}
	\label{hyper} 
\end{figure*}

\noindent\textbf{Ablation study} 
We also compare our LEF with or without attention layers. From the experiments, we can see that the extra attention layers will costs more runtime, while obviously achieving a better objective. The negative objective gaps refer to a better objective. Interestingly, for all the large-sized Datasets \uppercase\expandafter{\romannumeral2}, we could even achieve a better objective than the expert method, $\ell_p$-box ADMM itself. And with the problem size $n$ increases, the runtime speedup is also increasing. And with our attention layers, the accuracy of datasets \uppercase\expandafter{\romannumeral2} can be greater than $99.8\%$, which is impressive. These results exhibit the efficiency of attention layers within our early fixing framework. We also evaluate the efficiency of weighted loss in training, as shown in Fig. \ref{hyper}(b). We set the fixing threshold to 0.9, where all the constraints are feasible. From the figures, we can tell that the weighted loss generally achieves a smaller objective gap and a larger runtime speedup, compared to the no weighted loss.  

% We compare our EF framework with ES. From the experiments, ES contains an explicit objective gap, generally performing worse than EF. 
% When applying EF to LR, both the objective accuracy and runtime achieve better results. When applying EF to $\ell_p$box, the runtime obtains dramatic speedup, with less than 5\% degradation in the final objective accuracy. Especially in the Extra Hard instances, for $\ell_p$box with EF and AM, the runtime is nearly 60 times faster than that of the $\ell_p$box, and LR is not only accelerated by at least 3 $\times$, but also get improved on the final objective on four problems by 7\%, 32\%, 24\% and 0.7\%, respectively.

% \noindent\textbf{Ablation study}
% To show the necessity of neural network in EF framework and the efficiency of AM, we compare AM with a simple MLP and our hand-designed function (HF). From table \ref{tab1}, we can read that such neural network as MLP and AM could generally outperform HF, which infers the necessity of neural network. And we may infer that neural network is indeed able to discover more hidden information from the given knowledge. The comparison between the complex neural network with AM and the simple neural network with MLP figures out the competitiveness of AM. They are quite similar in runtime, with MLP a little bit faster due to the simplification of network structure. And AM is obviously more accurate than MLP in terms of the objective effectiveness.  

\noindent\textbf{Hyperparameter study} 
We analyze one of the most important hyperparameters in our early fixing framework: the fixing threshold $\theta$, as shown in Fig. \ref{hyper}(a). We evaluate how the number of infeasible constraints go with the fixing threshold, for different problem sizes in dataset \uppercase\expandafter{\romannumeral2}. We set fixing threshold from 0.5 to 0.9. When $\theta$ is 0.5, all the variables are fixed after the first block of iterations. When $\theta$ is 1, no variable is early fixed at all. From the results, only when $\theta$ is greater than 0.9, the number of infeasible constraints is 0, which means that the solution with early fixing is feasible. 
% And we can also see that a smaller $\theta$ generally achieve slightly larger runtime speedup.

% We can infer from the figure that the overall best performance occurs when $C$ is near to $[0.8, 0.9]$, with low objective gap as well as low runtime. As for block size $b$, we try 25, 50, 100, 250 and 500 in the setting, with $C=0.8$. We can see that as the block size goes larger, the objective gap is gradually decreasing, while the runtime is increasing. Generally, the best performance comes at $b\in[50,100]$. More details of the hyperparameter experiments can be found in Appendix. 

% Approximate methods have shown promising performance on both effectiveness and efficiency for solving the IP problem. However, we observed that many approximate methods converge at sublinear rates, \ie, a large fraction of variables fluctuate around their final converged discrete states in very long iterations. 
% Inspired by this observation, in this paper, we propose an early fixing framework, different from and more intelligent than early stopping, which aims to accelerate the approximate method by early fixing these fluctuated variables to their converged states, while not significantly harming the converged performance. To the best of our knowledge, we are the first to propose the idea of early fixing. 
% In each block of iterations, a policy network will evaluate the posterior probability of each unfixed variable with respect to all candidate discrete states. 

\begin{table*}[!t]
\label{tab3}
    \centering
\caption{Performance evaluations for image segmentation on the PASCAL Visual Object Classes Challenge 2012 datasets (VOC2012).} %, training on 12800 instances and test on 1000 instances. The method RL+LSTM is by \cite{pan2020solving}
\label{tab3}
\resizebox{17.9cm}{!}{
\begin{tabular}{lccccccccc}
    \toprule[1.5pt]
     Size $\rightarrow$ && \multicolumn{2}{c}{$n=1e4$} & \multicolumn{2}{c}{$n=5e4$} & \multicolumn{2}{c}{$n=1e5$} & \multicolumn{2}{c}{$n=5e5$}  \\
     \cmidrule(lr){3-4}\cmidrule(lr){5-6}\cmidrule(lr){7-8}\cmidrule(lr){9-10}
     Model  & Lang. & Energy$\downarrow$ & Time$\downarrow$ & Energy$\downarrow$ & Time$\downarrow$ & Energy$\downarrow$ & Time$\downarrow$ & Energy$\downarrow$ & Time$\downarrow$\\
    % \midrule
    \cmidrule(lr){1-2}\cmidrule(lr){3-4}\cmidrule(lr){5-6}\cmidrule(lr){7-8}\cmidrule(lr){9-10}
    min-cut \cite{boykov2004experimental} & M. & $\textbf{8778}$ & $\textbf{0.1s}$ & $\textbf{41049}$ & $\textbf{0.2s}$ & $\textbf{79737}$ & 0.4s & $\textbf{378439}$ & 1.8s  \\
    spectral relaxation \cite{shi2000normalized}  & M. & 10753 & 0.1s & 51192 & 0.2s & 105528 & $\textbf{0.3s}$ & 448737 & $\textbf{1.1s}$   \\
    linear relaxation \cite{dantzig2016linear} & M. & 9157 & 1.0s & 42275 & 5.5s & 81631 & 12.1s & 472253 & 121.0s     \\
    % $\ell_{1}$-box ADMM & M. & 12200 & 1.9s & 54581 & 7.2s & 105905 & 17.5s & 495638 & 99.8s  \\ 
    $\ell_{p}$-box ADMM \cite{wu2019lp} & M. & 8864 & 1.8s & 41181 & 7.1s & 79897 & 17.8s & 379860 & 98.9s   \\
    % \midrule
    \cmidrule(lr){1-2}\cmidrule(lr){3-4}\cmidrule(lr){5-6}\cmidrule(lr){7-8}\cmidrule(lr){9-10}
    % $\ell_{1}$-box ADMM & C.   \\
     && Energy$\downarrow$ \ Gap$\downarrow$ & Time$\downarrow$ & Energy$\downarrow$ \ Gap$\downarrow$ & Time$\downarrow$ & Energy$\downarrow$ \ Gap$\downarrow$ & Time$\downarrow$  & Energy$\downarrow$ \ Gap$\downarrow$ & Time$\downarrow$ \\ 
    \cmidrule(lr){3-4}\cmidrule(lr){5-6}\cmidrule(lr){7-8}\cmidrule(lr){9-10}
    $\ell_{p}$-box ADMM & C.+P. & $\textbf{8864}$ \ N/A & 1.3s & $\textbf{41181}$ \ N/A & 1.4s & $\textbf{79897}$ \ N/A & 3.1s & $\textbf{379860}$ \ N/A  & 20.2s  \\
    % $\ell_{p}$-box ADMM + HEF & C.+P.   \\
    $\ell_{p}$-box ADMM + LEF(w/o Att.) & C.+P. & 9341 \ 6.7\% & $\textbf{0.1s}$ & 43168 \ 6.4\% & $\textbf{0.5s}$ & 83612 \ 6.5\% & $\textbf{1.0s}$ & 395511 \ 6.1\% & $\textbf{5.0s}$  \\  
    $\ell_{p}$-box ADMM + LEF(with Att.) & C.+P.& 9124 \  $\textbf{3.5\%}$ & 0.1s & 42334 \ $\textbf{3.6\%}$ & 0.5s & 82009 \ $\textbf{3.5\%}$ & 1.1s & 388722 \ $\textbf{3.4\%}$ & 5.5s  \\   
    \cmidrule(lr){1-2}\cmidrule(lr){3-4}\cmidrule(lr){5-6}\cmidrule(lr){7-8}\cmidrule(lr){9-10}
    % \\
    
    Size $\rightarrow$ && \multicolumn{2}{c}{$n=1e6$} & \multicolumn{2}{c}{$n=5e6$} & \multicolumn{2}{c}{$n=1e7$} & \multicolumn{2}{c}{$n=5e7$} \\
    \cmidrule(lr){3-4}\cmidrule(lr){5-6}\cmidrule(lr){7-8}\cmidrule(lr){9-10}
    Model  & Lang. & Energy$\downarrow$ & Time$\downarrow$ & Energy$\downarrow$ & Time$\downarrow$ & Energy$\downarrow$ & Time$\downarrow$ & Energy$\downarrow$ & Time$\downarrow$\\
    % \midrule
    \cmidrule(lr){1-2}\cmidrule(lr){3-4}\cmidrule(lr){5-6}\cmidrule(lr){7-8}\cmidrule(lr){9-10}
    min-cut \cite{boykov2004experimental} & M. & $\textbf{741741}$ & 3.7s & $\textbf{4036901}$ & 22.8s & $\textbf{8006436}$ & 46.3s & $\textbf{48931748}$ & $\textbf{952.0s}$  \\
    spectral relaxation \cite{shi2000normalized} & M. & 1075301 & $\textbf{2.4s}$ & 4248567 & $\textbf{11.6s}$ & 8263794 & $\textbf{25.6s}$ & 57834141 & 986.9s \\
    linear relaxation \cite{dantzig2016linear} & M. & 883686 & 482.9s & 4405514 & 1152.5s & N/A & 3600s & N/A & 3600s      \\
    % $\ell_{1}$-box ADMM & M. & 957234 & 187.4s & 4440105 & 1110.8s & 8621096 &  2183.1s \\ 
    $\ell_{p}$-box ADMM \cite{wu2019lp} & M. & 744442 & 170.3s & 4056762 & 1170.2s & 8037384 &  2183.1s & N/A & 3600s \\
    % \midrule
    \cmidrule(lr){1-2}\cmidrule(lr){3-4}\cmidrule(lr){5-6}\cmidrule(lr){7-8}\cmidrule(lr){9-10}
    % $\ell_{1}$-box ADMM & C.   \\
     && Energy$\downarrow$ \ Gap$\downarrow$ & Time$\downarrow$ & Energy$\downarrow$ \ Gap$\downarrow$ & Time$\downarrow$ & Energy$\downarrow$ \ Gap$\downarrow$ & Time$\downarrow$  & Energy$\downarrow$ \ Gap$\downarrow$ & Time$\downarrow$ \\ 
    \cmidrule(lr){3-4}\cmidrule(lr){5-6}\cmidrule(lr){7-8}\cmidrule(lr){9-10}
    $\ell_{p}$-box ADMM \cite{wu2019lp} & C.+P. & $\textbf{744442}$ \ N/A & 41.9s & $\textbf{4056762}$ \ N/A & 293.6s & $\textbf{8037384}$ \ N/A & 548.1s & $\textbf{49075089}$ \ N/A & 3143.1s   \\
    % $\ell_{p}$-box ADMM + HEF & C.+P.   \\
    $\ell_{p}$-box ADMM + LEF(w/o Att.) & C.+P. & 773693 \ 5.9\% & $\textbf{10.6s}$ & 4191890 \ 4.9\% & $\textbf{56.7s}$ & 8289911 \ 4.7\% & $\textbf{144.5s}$ &  49339894 \ 4.5\% & $\textbf{981.1s}$ \\ 
    $\ell_{p}$-box ADMM + LEF(with Att.) & C.+P. & 760334 \ $\textbf{3.1\%}$ & 11.3s & 4123729 \ $\textbf{2.1\%}$ & 59.9s & 8153911 \ $\textbf{1.9\%}$ & 145.4s & 49286140 \ $\textbf{1.6\%}$ & 1108.14s  \\
    \bottomrule[1.5pt]
\end{tabular}}
\end{table*}

\section{MRF energy minimization}
\label{mrf}

\subsection{Formulations}
We consider the pairwise Markov Random Field (MRF) energy minimization problem based on a graph, which can be generally formulated as \cite{koller2009probabilistic}: 
\begin{align}
    \label{eq17}
  \mathop{\arg\min}_{\textbf{x}} \ \mathbb{E}(\textbf{x}) = \textbf{x}^\top\textbf{A}\textbf{x} +  \textbf{b}^\top\textbf{x}, \\  
  s.t. \ \ \textbf{Cx} {=} \textbf{1}, \ \ \textbf{x}{\in}\{0,1\}^{nK\times 1}. \nonumber 
\end{align}

\noindent where $\textbf{x}$ is a concatenation of all indicator vectors for the states $\kappa \in \{1,...,K\}$ and all $n$ nodes. If $x_{\kappa}^i = 1$, then node $i$ is on the state $\kappa$; otherwise, $x_{\kappa}^i = 0$. Each node can only take on one state, therefore we ensure that $\sum_{\kappa=1}^K x_{\kappa}^i = \textbf{1}$ for $\forall i, i \in \{1,...,n\}$. Thus we have the constraint set as in Formulation \ref{eq17}. As for the objective function, $\textbf{A}$ is the un-normalized Laplacian of the graph, $\textbf{A} = \textbf{D} - \textbf{W}$, $\textbf{W}$ is the matrix of node-to-node similarities. When $K=2$, the problem turns to a submodular minimization problem and it can be globally optimized using the min-cut algorithm \cite{boykov2004experimental} in polynomial time, however, it cannot be guaranteed when $K>2$.

% f1;...;Kg and all n nodes. For example, if xik 1⁄4 1, thennode i takes on the state k; otherwise, xik 1⁄4 0. Since each node
% can only take on one state, we enforce that PKk1⁄41 xik 1⁄4 1 for
% 8i 2 V, which is formulated as a sparse linear system of equal-
% ities: C x 1⁄4 1. Here, L 2 RnKnK is the un-normalized Lapla- 1
% cian of G, i.e., L 1⁄4 D  W with W being the matrix of node-to- node similarities. Our ‘p-box ADMM algorithm in Section 4 can be used to solve Eq. (24). Interestingly, practical graph constraints can be embedded into Eq. (24) as linear con- straints, such as hard (some nodes should have a particular state), mutually exclusive (some nodes should have different states) and cardinality (a limit on the number of nodes belong- ing to a particular state) constraints.
% Popular Methods. It has been proven that when K 1⁄4 2, Eq. (24) is a submodular minimization problem, and it can be globally optimized using the min-cut algorithm (a discrete method) in polynomial time [21], [46]. However, when K > 2, this global solution cannot be guaranteed in general.

\subsection{Experiments for Image segmentation}

\noindent\textbf{Accelerated method and datasets} $\ell_p$-Box ADMM \cite{wu2019lp} has been proved to be efficient in image segmentation when $K=2$. Thus in our experiment, we choose $\ell_p$-Box ADMM to be accelerated. We also regard the min-cut \cite{boykov2004experimental} as the ground-truth algorithm ($K{=}2$) for comparisons. The PASCAL Visual Object Classes Challenge 2012 (VOC2012) dataset \cite{everingham2015pascal} has been widely used in computer vision tasks, such as object classification, object detection and object segmentation. We thus choose VOC2012 for our experiments, where 2913 images are available for segmentation. We then randomly select 100, 20, 20 images for the training, validation and testing, respectively. We resize the testing images to different sizes and do the testing, including $n=1e4, 5e4, 1e5, 5e5, 1e6, 5e6, 1e7$ and $5e7$.   

\noindent\textbf{Training details} We train our model only on the smallest sized images with $n=1e4$, and generalize to all other larger-sized instances. $\gamma, \beta, \delta, \mathbb{L}$ are set to 5, 10, 0.9, 2, respectively. The learning rate is set to $1e{-}4$.  We train for 20 epochs. For our learning-based early fixing methods, training without attention for one epoch costs 122s, while training with attention costs 129s. We implement the functions of the accelerated methods in C++ and call the functions in Python via Cython interfaces. All the learning modules are implemented in Python.

\noindent\textbf{Baselines} We compare our method against two generic IP solvers, namely linear relaxation \cite{dantzig2016linear} and spectral relaxation \cite{shi2000normalized}, and a state-of-the-art and widely used min-cut algorithm \cite{boykov2004experimental}. The linear relaxation is implemented using the built-in function \textit{quadprog} in MATLAB. The closed-form solution based on eigen-decomposition is implemented for spectral relaxation. We also compare against the $\ell_p$-box ADMM in both MATLAB and C++. We then apply our learning-based early fixing framework to $\ell_p$-box ADMM, with or without attention layers. We implement the functions of the accelerated methods in C++ and call the functions in Python via Cython interfaces.

\noindent\textbf{Evaluations} We evaluate all the different sizes, each with 20 instances. We evaluate the energy and the runtime for MATLAB implementations, and for our early fixing we also evaluate the energy gap towards the accelerated method itself. We report the mean values for all the 20 instances of different sizes. The time limit is set to 1 hour.

\noindent\textbf{Comparison study} The experimental results are shown in Table \ref{tab3}. We compare all methods in terms of their final energy value and runtime in the case of binary submodular MRF ($K$=2). From the results, we can see that min-cut achieves the lowest energy, regarded as the ground-truth methods for the $K$=2 image segmentation task. Among all the methods, spectral relaxation achieves the worst performance, though it is fast in computation. Linear relaxation runs fast in small sized instances, while running slow for large scale instances. Linear relaxation obtains smaller energy than spectral relaxation, while it is much worse than $\ell_p$-box ADMM. The C++ implementation of $\ell_p$-box ADMM runs much faster than MATLAB version with significant speedup. Compared to the accelerated $\ell_p$-box ADMM, our early fixing framework generally achieves from $3\times$ to $5\times$ speedup in runtime. Especially for the smallest sized instances with $n=1e4$, the runtime speedup is even greater than $10 \times$. In Fig. \ref{testing_seg}, we record how the energy changes with respect to the iterations during the inference stages for one image in different sizes. From the figure, we can see that our LEF method leads to much faster convergence, and LEF with attention layers achieves lower energy.

\noindent\textbf{Ablation study} We also show the ablation study of our learning-based early fixing. We compare the results with or without attention layers. Equipping with the attention layers will cost a little bit more runtime, however, the energy will decrease a bit more. For LEF without attention layers, the average energy gap of all different sizes is 5.7\%, while the average energy gap with attention layers is only 2.8\%. The gap is overall decreased by 2.9\%. 

\begin{figure*}[!htb] %H为当前位置，!htb为忽略美学标准，htbp为浮动图形
  \centering %图片居中
  \includegraphics[width=1.0\textwidth]{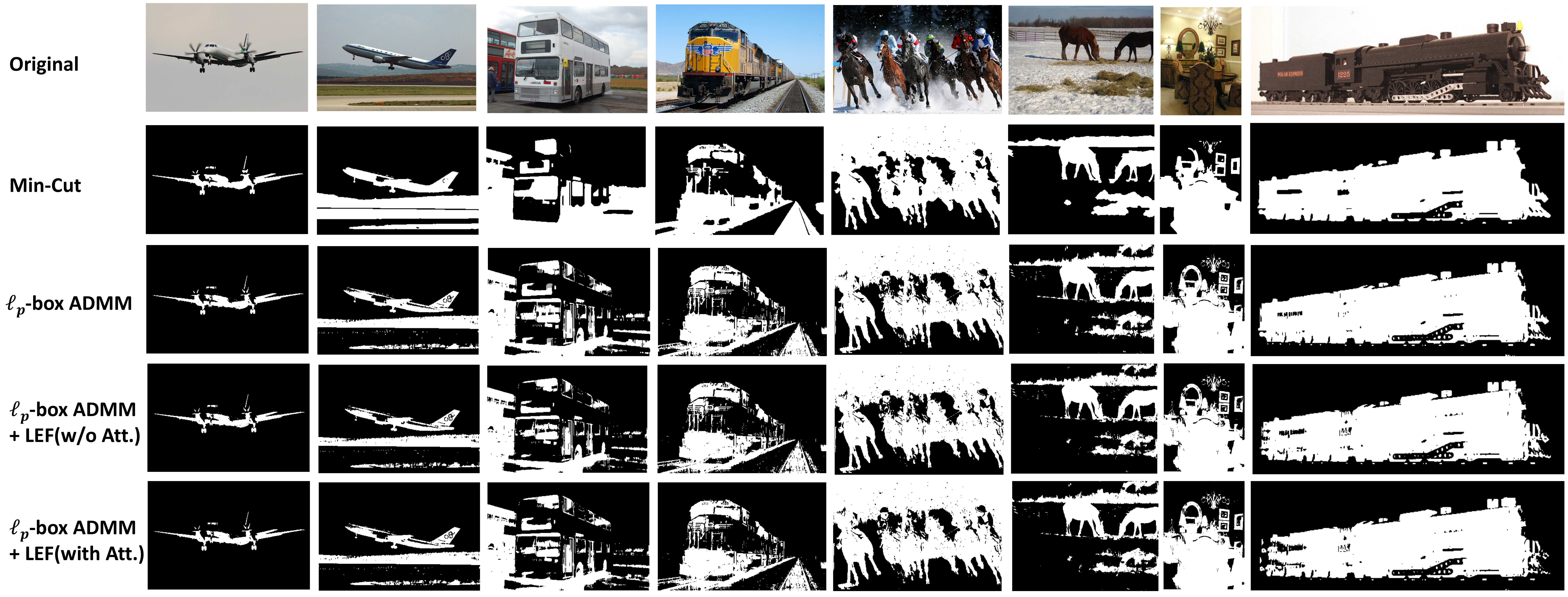} %插入图片，[]中设置图片大小，{}中是图片文件名
  \caption{Performance exhibition of different methods for image segmentation on the PASCAL Visual Object Classes Challenge 2012 datasets (VOC2012). $n{=}1e5$. Min-cut can obtain optimal solutions when $K{=}2$. $\ell_p$-box ADMM is the method to be accelerated.} %最终文档中希望显示的图片标题
  \label{fig_seg} %用于文内引用的标签
\end{figure*}

\begin{figure*}[!htb] %H为当前位置，!htb为忽略美学标准，htbp为浮动图形
    \centering %图片居中
    \includegraphics[width = 0.99\textwidth]{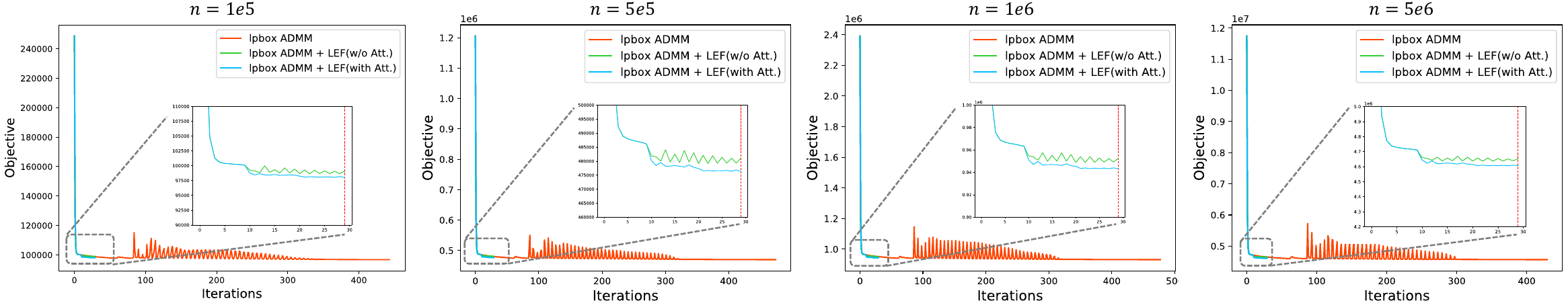}  
	\caption{Convergence on different methods for image segmentation: how the objective changes with respect to the iterations during the inference stages for one images in four different sizes. LEF method leads to much faster convergence, and LEF with attention layers achieves lower energy.}
	\label{testing_seg} 
\end{figure*}

\noindent\textbf{Performance exhibition} We exhibit some segmented images of different shapes by the optimal min-cut algorithm, $\ell_p$-box ADMM as well as our learning-based early fixing with or without attention layers. From Figure \ref{fig_seg}, we can see that $\ell_p$-box ADMM generally achieves a great segmentation performance. And with our early fixing framework, the segmentation efficiency is also excellent. 
% Visually, it is hard to tell the different between images with LEF and without LEF. 


\section{Sparse adversarial attack}
\label{saa}

\subsection{Background}
We consider the sparse adversarial attack \cite{fan2020sparse}, which generates adversarial perturbations onto partial positions of the clean image, where the perturbed image is incorrectly predicted by the deep model. There are two challenges lying in the sparse adversarial attack. One is where to perturb and the other is how to determine the perturbation magnitude. Some works manually or heuristically determined the perturbed positions, and optimized the magnitude using an appropriate algorithm designed for the dense adversarial attack. However, Fan et. al. \cite{fan2020sparse} proposed to factorize the perturbation at each pixel to the product of two variables, including the perturbation magnitude and one binary selection factor (i.e., 0 or 1). One pixel is perturbed if its selection factor is 1, otherwise not perturbed. The perturbation $\bm{\epsilon}$ can be factorized as: 
\begin{align}
	\label{eq18}
	\bm{\epsilon} = \bm{\zeta} \odot \bm{\eta},
\end{align}
where $\bm{\zeta} \in \mathbb{R}^n$ denotes the vector of perturbation magnitudes, and $\bm{\eta} \in \{0,1\}^n$ denotes the vector of perturbed positions. $\odot$ represents the element-wise product. Then the sparse attack problem can be formulated as a mixed integer programming (MIP) by jointly optimizing the continuous perturbation magnitudes $\bm{\zeta}$ and the binary selection factors $\bm{\eta}$ of all pixels. Inspired by $\ell_p$-box ADMM \cite{wu2019lp}, they proposed to reformulate the MIP problem to an equivalent continuous optimization problem. They update the $\bm{\zeta}$ by gradient descent, and update the $\bm{\eta}$ by ADMM. At this point, we are going to accelerate the $\bm{\eta}$ updating parts with our early fixing framework. 

% Besides, the perturbation factorization provides the extra flexibility to incorporate other meaningful constraints on selection factors or magnitudes to achieve some desired performance, such as the group-wise sparsity or the enhanced visual imperceptibility. We develop an efficient algorithm by equivalently reformulating the MIP problem as a continuous optimization problem. Extensive experiments demonstrate the superior- ity of the proposed method over several state-of-the-art sparse attack methods.

\subsection{Adversarial attack experiments}

% \begin{figure*}[!t] %H为当前位置，!htb为忽略美学标准，htbp为浮动图形
%     \centering %图片居中
%     \includegraphics[width = 0.98\textwidth]{IEEEtran/figures/advAtt.png} 
% 	\caption{Examples of perturbations generated by the proposed SAPF method. (Left- most column): Two benign images from ImageNet with their ground-truth class labels given below; (2nd - 5th columns): the generated sparse adversarial perturbations with different target attack classes given below; (Right-most column): the common perturbed pixels of four targeted adversarial attacks}
% 	\label{adv} 
% \end{figure*}

\textbf{Datasets and models} We follow the setup in SAPF (Sparse adversarial Attack via Perturbation Factorization) \cite{fan2020sparse}, and use the CIFAR-10 \cite{krizhevsky2009learning} and ImageNet \cite{deng2009imagenet} for the experiments. There are 50k training images and 10k validation images covering 10 classes for CIFAR-10. We randomly select 1000 images from the validation set for our experiments. Each image has 9 target classes except its ground-truth class. Thus there are totally 9000 adversarial examples for the adversarial attack method. ImageNet contains 1000 classes, with 1.28 million images for training and 50k images for validation. We randomly choose 100 images covering 100 different classes from the validation set. To reduce the time complexity, we randomly select 9 target classes for each image in ImageNet, resulting in 900 adversarial examples. As regard to the classification model, on CIFAR-10, we follow the setting of C\&W \cite{carlini2017towards} and train a network that consists of four convolution layers, three fully-connected layers, and two max-pooling layers. The input size of the network is 32 x 32 x 3. On ImageNet, we use a pre-trained Inception-v3 network \cite{szegedy2016rethinking}. The input size of the network is 299 x 299 x 3. All other experimental hyper-parameters for SAPF are set to default \cite{fan2020sparse}.  

\begin{table*}[!t]
    \centering
\caption{Performance comparison of targeted sparse adversarial attack on CIFAR-10 and ImageNet. We consider the ASR and $\ell_p$-norm ($p$ = 0, 1, 2, ${\infty}$) of the learned perturbation.} %, training on 12800 instances and test on 1000 instances. The method RL+LSTM is by \cite{pan2020solving}
\label{tab5}
\resizebox{17.9cm}{!}{
\begin{tabular}{llccccccccccccccc}
    \toprule[1.5pt]
     \multirow{2}{*}{Dataset} & \multirow{2}{*}{Method} & \multicolumn{5}{c}{Best case} & \multicolumn{5}{c}{Average case} & \multicolumn{5}{c}{Worse case} \\
      &  & ASR(\%) & $\ell_{0}$ & $\ell_{1}$ & $\ell_{2}$ & $\ell_{\infty}$ & ASR(\%) & $\ell_{0}$ & $\ell_{1}$ & $\ell_{2}$ & $\ell_{\infty}$ & ASR(\%) & $\ell_{0}$ & $\ell_{1}$ & $\ell_{2}$ & $\ell_{\infty}$ \\
    % \midrule
    \cmidrule(lr){1-2}\cmidrule(lr){3-7}\cmidrule(lr){8-12}\cmidrule(lr){13-17}
    \multirow{8}{*}{CIFAR-10} & One-pixel \cite{su2019one} &  15.0  & 3 & 1.57 & 0.96 & 0.68 & 5.5 & 3 & 2.19 & 1.29 & 0.82 & 0.7 & 3 & 2.66 & 1.54 & 0.92 \\
    & CornerSearch \cite{croce2019sparse} & 60.4 & 537 & 69.70 & 3.34 & 0.34 & 59.3 & 549 & 73.64 & 3.48 & 0.34 & 63.2 & 77.57 & 561 & 3.62 & 0.34  \\
    & PGD $\ell_0$+$\ell_{\infty}$ \cite{croce2019sparse} & 99.4 & 555 & 18.11 & 0.97 & 0.12 & 98.6 & 555 & 23.17 & 1.17 & 0.12 & 99.3 & 555 & 26.82 & 1.35 & 0.13  \\
    & SparseFool \cite{modas2019sparsefool} & 100 & 255 & 11.87  & 0.67 & 0.05 & 99.9 & 555 & 25.81 & 1.04 & $\textbf{0.05}$ & 99.8 & 852 & 39.67 & 1.34 & $\textbf{0.05}$   \\ 
    & C\&W-$\ell_0$ \cite{carlini2017towards} &  100 & 614 & 6.95 & 0.43 & 0.09 & 100 & 603 & 13.07 & 0.81 & 0.16 & 100 & 598 & 18.60 & 1.14 & 0.22  \\
    & StrAttack \cite{xu2018structured} & 100 & 391 & 4.94 & 0.30 & 0.05 & 100 & 543 & 9.49 & 0.52 & 0.09 & 100 & 476 & 12.44 & 0.77 & 0.14 \\ 
    & SAPF \cite{fan2020sparse} & 100 & 387 & 4.61 & 0.25 & 0.04 & 100 & 603 & 8.51 & 0.44 & 0.06 & 100 & 471 & 10.39 & 0.60 & 0.10  \\
    & SAPF + LEF(w/o Att.) & 100 & $\textbf{149}$ & 5.23 & 0.46 & 0.10 & 100 & $\textbf{303}$ & 8.48 & 0.64 & 0.10 & 100 & $\textbf{459}$ & 11.69 & 0.62 & 0.10  \\
    & SAPF + LEF(with Att.) & $\textbf{100}$ & 276 & $\textbf{4.43}$ & $\textbf{0.25}$ & $\textbf{0.04}$ & $\textbf{100}$ & 510 & $\textbf{8.37}$ & $\textbf{0.44}$ & 0.06 & $\textbf{100}$ & 506 & $\textbf{10.24}$ & $\textbf{0.55}$ & 0.09  \\
    
    \cmidrule(lr){1-2}\cmidrule(lr){3-7}\cmidrule(lr){8-12}\cmidrule(lr){13-17}
    \multirow{8}{*}{ImageNet} & One-pixel \cite{su2019one} & 0 & 3 & 1.19 & 0.80 & 0.66 & 0 & 3 & 1.88 & 1.18 & 0.83 & 0 & 3 & 2.56 & 1.51 & 0.93  \\
    & CornerSearch \cite{croce2019sparse} & 4 & 58658 & 5962.46 & 28.06 & 0.44 & 1.3 & 58792 & 6018.31 & 28.29 & 0.44 & 2 & 58920 &6076.07 & 28.53 & 0.44   \\
    & PGD $\ell_0$+$\ell_{\infty}$ \cite{croce2019sparse} & 95 & 56922 & 798.89 & 4.21 & 0.06 & 95.6 & 56919 & 854.67 & 4.51 & 0.06 & 96 & 56920 & 925.27 & 4.90 & 0.06 \\
    & SparseFool \cite{modas2019sparsefool} & 97 & 34205 & 174.17 & 0.92 & $\textbf{0.01}$ & 80.6 & 59940 & 305.18 & 1.22 & $\textbf{0.01}$ & 46 & 82576 & 420.44 & 1.45 & $\textbf{0.01}$ \\ 
    & C\&W-$\ell_0$ \cite{carlini2017towards} & 100 & 73407 & 133.79 & 0.79 & 0.05 & 100 & 70885 & 199.20 & 1.12 & 0.06 & 100 & 69947 & 269.10 & 1.46 & 0.07 \\
    & StrAttack \cite{xu2018structured} & 100 & 38354 & 77.28 & 0.69 & 0.06 & 100 & 58581 & 127.59 & 0.97 & 0.08 & 100 & 67348 & 171.25 & 1.28 & 0.10 \\ 
    & SAPF \cite{fan2020sparse} & 100 & 37275 & 70.25 & $\textbf{0.59}$ & 0.04 & 100 & 56218 & 112.16 & $\textbf{0.72}$ & 0.04 & 100 & 56843 & 150.55 & $\textbf{1.12}$ & 0.04 \\
    & SAPF + LEF(w/o Att.) & 100 & $\textbf{4146}$ & 54.21 & 1.19 & 0.10 & 100 & $\textbf{4074}$ & 78.32 & 1.36 & 0.10 & 100 & $\textbf{4570}$ & 111.09 & 1.75 & 0.10\\
    & SAPF + LEF(with Att.) & $\textbf{100}$ & 5311 & $\textbf{47.44}$ & 0.95 & 0.08 & $\textbf{100}$ & 5344 & $\textbf{67.26}$ & 1.08 & 0.08 & $\textbf{100}$ & 5582 & $\textbf{88.16}$ & 1.28 & 0.08 \\
    
    \bottomrule[1.5pt]
\end{tabular}}
\end{table*}

% $\textbf{}$
\begin{table}[!t]
    \centering
\caption{Runtime comparison of targeted sparse adversarial attack on CIFAR-10 and ImageNet. We record the runtime for $\bm{\zeta}$ updating, $\bm{\eta}$ updating, and total. The fourth column is the time speedup for $\bm{\zeta}$ updating which uses our early fixing to accelerate $\ell_p$-box ADMM.}
\label{tab6}
\resizebox{8.8cm}{!}{
\begin{tabular}{llccccc}
    \toprule[1.5pt]
     {Dataset} & {Method} & $\bm{\eta}$ Updating & $\bm{\eta}$ Speedup & $\bm{\zeta}$ Updating & Total \\
    
    \cmidrule(lr){1-2}\cmidrule(lr){3-6}
    \multirow{3}{*}{CIFAR-10} & SAPF \cite{fan2020sparse} & 79.6s & N/A & 81.3s & 160.8s \\
    & SAPF + LEF(w/o Att.) & $\textbf{14.1s}$ & $\textbf{5.6}\times$ & 81.5s & 95.6s  \\
    & SAPF + LEF(with Att.) & 14.4s & 5.5$\times$ & $\textbf{79.7s}$ & $\textbf{94.1s}$ \\
    
    \cmidrule(lr){1-2}\cmidrule(lr){3-6}
    \multirow{3}{*}{ImageNet} & SAPF \cite{fan2020sparse} & 499.2s & N/A & $\textbf{560.1s}$ & 1059.3s  \\
    & SAPF + LEF(w/o Att.) & $\textbf{204.4s}$ & $\textbf{2.4}\times$ & 561.5s & $\textbf{765.9s}$  \\
    & SAPF + LEF(with Att.) & 231.2s & 2.2$\times$ & 565.3s & 796.5s  \\

    \bottomrule[1.5pt]
\end{tabular}}
\end{table}
\begin{figure}[!t] %H为当前位置，!htb为忽略美学标准，htbp为浮动图形
    \centering %图片居中
    \includegraphics[width = 0.48\textwidth]{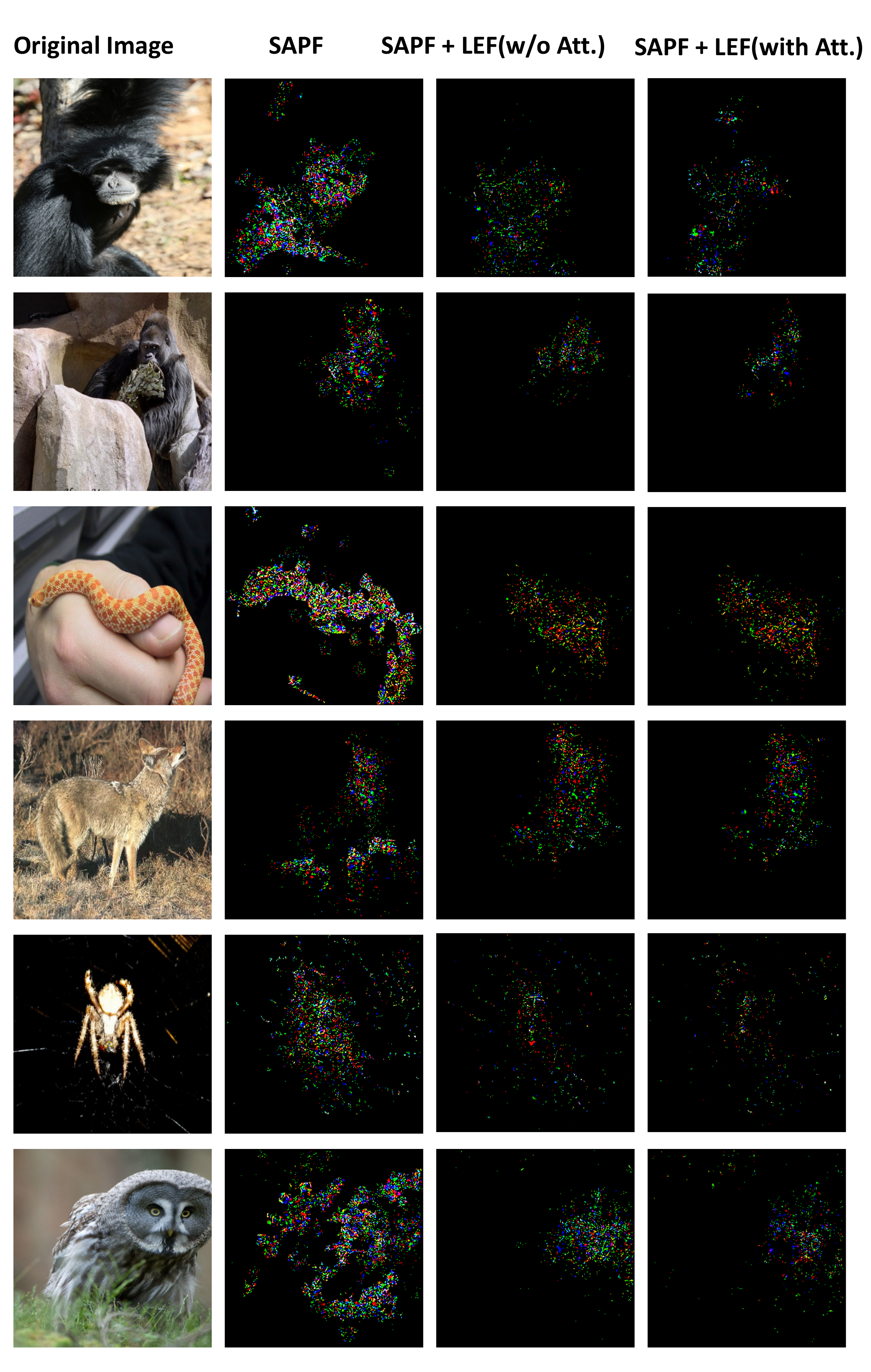} 
	\caption{Examples of perturbations generated by the SAPF method, the SAPF with our early fixing framework, equipped without or with attention layers. From the top to the bottom, the ground-truth class and target class label pairs are: (siamang, zucchini), (gorilla, Brabancon griffon), (hognose snake, red-backed sandpiper), (coyote, impala), (barn spider, greenhouse), (great grey owl, capuchin).
}
	\label{sparse} 
\end{figure}

\noindent\textbf{Training details} As regard to our early fixing framework, we randomly pick 20 images from CIFAR-10. And each image corresponds to one MIP problem. By default, solving one MIP problem for adversarial attack is with 6 search loops for $\bm{G}$ updating. Thus we record these 20*6=120 instances for early fixing training. $\gamma, \beta, \delta, \mathbb{L}$ are set to 3, 50, 0.9, 2, respectively. The learning rate is set to $1e{-}4$.  We train for 20 epochs. For our learning-based early fixing methods, training without attention for one epoch costs 4min 28s, while training with attention costs 4min 20s. All the implementations are based on Python. We use the pre-trained model on CIFAR-10 to test on both CIFAR-10 and ImageNet. 

\noindent\textbf{Baselines and evaluations} We compare whether to use our learning-based early fixing (LEF) on SAPF or not. We also compare whether to use attention layers or not. Besides, we also record the results by other attack paradigms, including one-pixel \cite{su2019one}, corner search \cite{croce2019sparse}, PGD $\ell_0$+$\ell_{\infty}$ \cite{croce2019sparse}, SparseFool \cite{modas2019sparsefool}, C\&W-$\ell_0$ \cite{carlini2017towards}, StrAttack \cite{xu2018structured}. Those results of other attacks are from the SAPF paper \cite{fan2020sparse}.

As regard to the evaluations, the $\ell_p$-norm ($p$ = 0, 1, 2, ${\infty}$) of perturbations and the attack success rate (ASR) are utilized to evaluate the attack performance of different methods. We follow the same setting in \cite{fan2020sparse}. We keep increasing the upper bound of $\ell_p$-norm of perturbations until the attack is success. We compare different attack algorithms in terms of the $\ell_p$-norm of perturbations under 100\% ASR, though some sparse attack methods fail to generate 100\% ASR. Moreover, for each image, we evaluate three different cases, i.e., the average case: the average performance of all 9 target classes; the best case: the performance w.r.t. the target class that is the easiest to attack; and the worst case: the performance w.r.t. the target class that is the most difficult to attack.

\noindent\textbf{Results analysis} We exhibit the attack performances of best/average/worst cases with $\ell_p$-norm and ASR in Table \ref{tab5}. We present the runtime comparisons whether to use LEF or not on SAPF in Table \ref{tab6}. We show the examples of generated perturbations by different methods in Figure \ref{sparse}. From the tables, we see that SAPF method achieve 100\% attack success rate under all three cases, in both CIFAR-10 and ImageNet datasets. And with our LEF, the ASR still remains 100\% in different cases, no matter with attention layers or not.   

The $\ell_{\infty}$-norm of the one-Pixel-Attack is the largest among all algorithms and it achieves the lowest attack success rate. It is difficult to perform targeted adversarial attacks by only perturbing one pixel (the $\ell_0$ = 3 relates to three channels), even on the tiny database CIFAR-10. The CornerSearch and PGD $\ell_0$+$\ell_{\infty}$ also fails to generate 100\% success attack rate. Comparing to all adversarial attack algorithms except one-Pixel-Attack, SAPF method achieves the best $\ell_1$-norm and $\ell_2$-norm under all three cases. On CIFAR-10, with our LEF with attention layers, it achieves the better $\ell_1$-norm and $\ell_2$-norm under all cases compared to SAPF. On ImageNet, it achieves the better $\ell_1$-norm under all cases. 

\begin{figure}[!t] %H为当前位置，!htb为忽略美学标准，htbp为浮动图形
    \centering %图片居中
    \includegraphics[width = 0.48\textwidth]{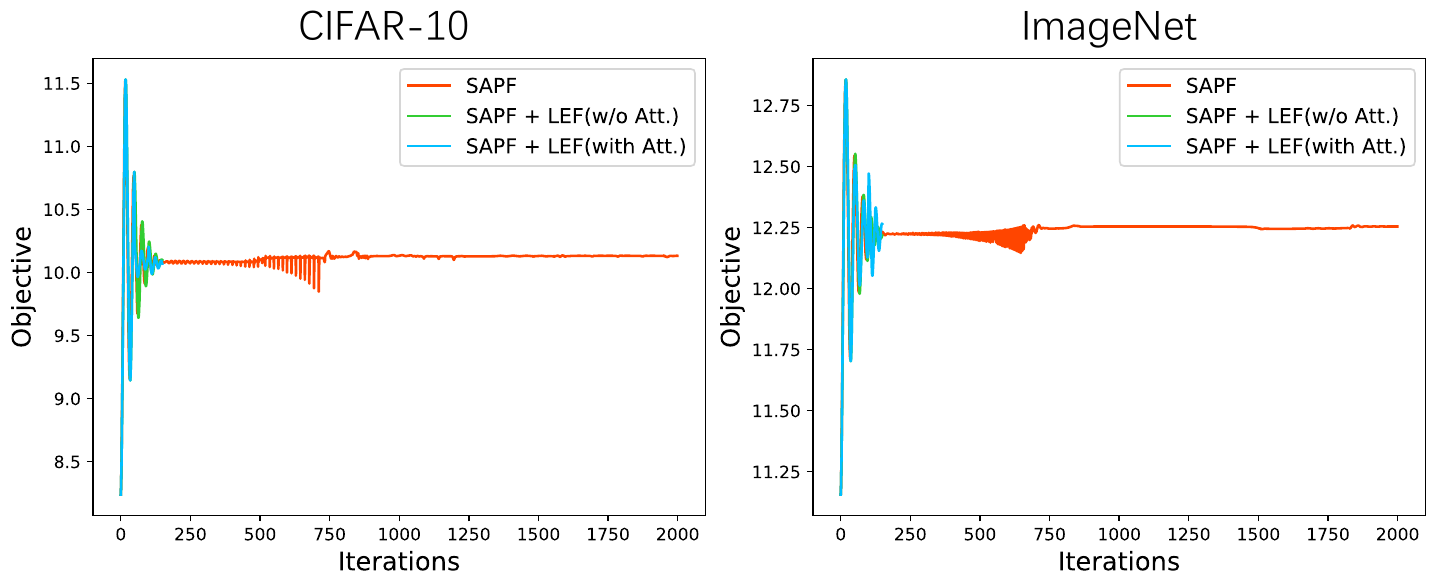}  
	\caption{Convergence on different methods for sparse adversarial attack: how the objective (loss) changes with respect to the iterations during the inference stages for one image in CIFAR-10 (left) and ImageNet (right).}
	\label{testing_sad} 
\end{figure}

More importantly, with the aim of algorithmic acceleration, we can see the obvious time speedup on the $\bm{\eta}$ updating part, by Table \ref{tab6}. On CIFAR-10 dataset, the time speedup is more than 5$\times$, while on ImageNet dataset, the time speedup is more than 2$\times$. Those results demonstrate the efficiency and effectiveness of the proposed LEF method. In Fig. \ref{testing_sad}, we record the first loop out of the 6 search loops for $\bm{\eta}$ Updating. The figure reveals that our LEF method leads to faster convergence than the SAPF itself, when they all have quite similar objective (loss).

% \section{Discussions}
% \label{conclusion}
% 1. Why only consider the iterative values as inputs: we also consider to use some constraints information as in the inputs, which could not be that general enough for all types of optimizations tasks, and this could be general enough for larger size problems. 2. How to decide the hyperparameters. And how they correlate with the early fixing performances.
% \noindent\textbf{Limitations}
% There are two limitations in our work. Firstly, there are so many training samples that the training process could be time-consuming. Since each variable is independent with one another, there could be $5000{\times}n$ training samples from 1,000 training instances. Taking the set covering problem where $n=500$ as an example, the training sample size is as large as $2.5$ millions. Secondly, there could be infeasibility issues in the early fixing process. If we early fix the variables to the wrong solutions, then in the later iterations, the whole optimization task may become infeasible, and the objective gap could possibly get enlarged. Thus we come up with the fixing confidence $C$, only when the output possibilities exceed the fixing confidence shall we fix the variables. From the results in figure \ref{hyper}, we can see that an appropriate fixing confidence helps alleviate the objective gap.   

\section{Conclusions and future work}
\label{conclusion}
We propose an early fixing framework, which aims to accelerate the approximate method by early fixing the fluctuated variables to their converged states, while not significantly harming the converged performance. To the best of our knowledge, we are the first to propose the framework of early fixing for solving integer programming. 
% A policy network will evaluate the posterior probability of each unfixed variable with respect to all discrete candidate states in each block of iterations.
We construct the whole early fixing process as a Markov decision process, and incorporate the imitation learning paradigm for training with the weighted binary cross-entropy loss. Specifically, we adopt the powerful attention layers in the policy network. 
We conduct the extensive experiments for our proposed early fixing framework to three different IP applications: constrained linear programming, MRF energy minimization and sparse adversarial attack. The experimental results in different scenarios demonstrate the efficiency of our proposed learning-based early fixing framework. 
In the future work, we would like to apply the early fixing framework to more approximate methods, and discover more possibilities to improve the efficiency and effectiveness of these methods. Meanwhile, there is a plenty of room to mitigate the objective gap when using some other efficient policy networks. 

% , for the largest scale instances, the proposed framework can accelerate the $\ell_p$-Box ADMM up to 60 $\times$, with less than $5\%$ degradation of the final objective effectiveness; the LR is not only accelerated by at least 3 $\times$, but also get improved on the final objective by 0.7\% - 32\%.

% \clearpage
% \medskip

% % use section* for acknowledgment
% \ifCLASSOPTIONcompsoc
%   % The Computer Society usually uses the plural form
%   \section*{Acknowledgments}
% \else
%   % regular IEEE prefers the singular form
% \fi


\ifCLASSOPTIONcaptionsoff
  \newpage
\fi




\bibliographystyle{unsrt}
\bibliography{ref}

\begin{thebibliography}{10}

\bibitem{khalil2016machine}
Elias~B Khalil.
\newblock Machine learning for integer programming.
\newblock In {\em IJCAI}, pages 4004--4005, 2016.

\bibitem{wang2014tracking}
Xinchao Wang, Engin T{\"u}retken, Fran{\c{c}}ois Fleuret, and Pascal Fua.
\newblock Tracking interacting objects optimally using integer programming.
\newblock In {\em European Conference on Computer Vision}, pages 17--32.
  Springer, 2014.

\bibitem{gunluk2021optimal}
Oktay G{\"u}nl{\"u}k, Jayant Kalagnanam, Minhan Li, Matt Menickelly, and Katya
  Scheinberg.
\newblock Optimal decision trees for categorical data via integer programming.
\newblock {\em Journal of Global Optimization}, pages 1--28, 2021.

\bibitem{lawler1966branch}
Eugene~L Lawler and David~E Wood.
\newblock Branch-and-bound methods: A survey.
\newblock {\em Operations research}, 14(4):699--719, 1966.

\bibitem{kelley1960cutting}
James~E Kelley, Jr.
\newblock The cutting-plane method for solving convex programs.
\newblock {\em Journal of the society for Industrial and Applied Mathematics},
  8(4):703--712, 1960.

\bibitem{mitchell2010branch}
John~E Mitchell.
\newblock Branch and cut.
\newblock {\em Wiley encyclopedia of operations research and management
  science}, 2010.

\bibitem{boyd2004convex}
Stephen Boyd and Lieven Vandenberghe.
\newblock {\em Convex optimization}.
\newblock Cambridge university press, 2004.

\bibitem{zha2001spectral}
Hongyuan Zha, Xiaofeng He, Chris Ding, Ming Gu, and Horst~D Simon.
\newblock Spectral relaxation for k-means clustering.
\newblock In {\em Advances in neural information processing systems}, pages
  1057--1064, 2001.

\bibitem{lasserre2001explicit}
Jean~B Lasserre.
\newblock An explicit exact sdp relaxation for nonlinear 0-1 programs.
\newblock In {\em International Conference on Integer Programming and
  Combinatorial Optimization}, pages 293--303. Springer, 2001.

\bibitem{boyd2011distributed}
Stephen Boyd, Neal Parikh, and Eric Chu.
\newblock {\em Distributed optimization and statistical learning via the
  alternating direction method of multipliers}.
\newblock Now Publishers Inc, 2011.

\bibitem{rantzer2009dynamic}
Anders Rantzer.
\newblock Dynamic dual decomposition for distributed control.
\newblock In {\em 2009 American Control Conference}, pages 884--888. IEEE,
  2009.

\bibitem{chatzipanagiotis2015augmented}
Nikolaos Chatzipanagiotis, Darinka Dentcheva, and Michael~M Zavlanos.
\newblock An augmented lagrangian method for distributed optimization.
\newblock {\em Mathematical Programming}, 152(1):405--434, 2015.

\bibitem{fu2013bethe}
Qiang Fu, Huahua Wang, and Arindam Banerjee.
\newblock Bethe-admm for tree decomposition based parallel map inference.
\newblock {\em arXiv preprint arXiv:1309.6829}, 2013.

\bibitem{wang2013bregman}
Huahua Wang and Arindam Banerjee.
\newblock Bregman alternating direction method of multipliers.
\newblock {\em The Conference on Neural Information Processing Systems}, 2014.

\bibitem{xie2019differentiable}
Xingyu Xie, Jianlong Wu, Guangcan Liu, Zhisheng Zhong, and Zhouchen Lin.
\newblock Differentiable linearized admm.
\newblock In {\em International Conference on Machine Learning}, pages
  6902--6911. PMLR, 2019.

\bibitem{liu2019linearized}
Qinghua Liu, Xinyue Shen, and Yuantao Gu.
\newblock Linearized admm for nonconvex nonsmooth optimization with convergence
  analysis.
\newblock {\em IEEE Access}, 7:76131--76144, 2019.

\bibitem{wu2019lp}
Baoyuan Wu and Bernard Ghanem.
\newblock $\ell p$-box admm: A versatile framework for integer programming.
\newblock {\em IEEE transactions on pattern analysis and machine intelligence},
  41(7):1695--1708, 2019.

\bibitem{zamir2017feedback}
Amir~R Zamir, Te-Lin Wu, Lin Sun, William~B Shen, Bertram~E Shi, Jitendra
  Malik, and Silvio Savarese.
\newblock Feedback networks.
\newblock In {\em Proceedings of the IEEE conference on computer vision and
  pattern recognition}, pages 1308--1317, 2017.

\bibitem{huang2017multi}
Gao Huang, Danlu Chen, Tianhong Li, Felix Wu, Laurens van~der Maaten, and
  Kilian~Q Weinberger.
\newblock Multi-scale dense networks for resource efficient image
  classification.
\newblock {\em arXiv preprint arXiv:1703.09844}, 2017.

\bibitem{vaswani2017attention}
Ashish Vaswani, Noam Shazeer, Niki Parmar, Jakob Uszkoreit, Llion Jones,
  Aidan~N Gomez, Lukasz Kaiser, and Illia Polosukhin.
\newblock Attention is all you need.
\newblock {\em arXiv preprint arXiv:1706.03762}, 2017.

\bibitem{howard1960dynamic}
Ronald~A Howard.
\newblock Dynamic programming and markov processes.
\newblock 1960.

\bibitem{torabi2018behavioral}
Faraz Torabi, Garrett Warnell, and Peter Stone.
\newblock Behavioral cloning from observation.
\newblock {\em arXiv preprint arXiv:1805.01954}, 2018.

\bibitem{wu2020map}
Baoyuan Wu, Li~Shen, Tong Zhang, and Bernard Ghanem.
\newblock Map inference via $\ell$2-sphere linear program reformulation.
\newblock {\em International Journal of Computer Vision}, 128(7):1913--1936,
  2020.

\bibitem{fan2020sparse}
Yanbo Fan, Baoyuan Wu, Tuanhui Li, Yong Zhang, Mingyang Li, Zhifeng Li, and
  Yujiu Yang.
\newblock Sparse adversarial attack via perturbation factorization.
\newblock In {\em European conference on computer vision}, pages 35--50.
  Springer, 2020.

\bibitem{li2019compressing}
Tuanhui Li, Baoyuan Wu, Yujiu Yang, Yanbo Fan, Yong Zhang, and Wei Liu.
\newblock Compressing convolutional neural networks via factorized
  convolutional filters.
\newblock In {\em Proceedings of the IEEE/CVF Conference on Computer Vision and
  Pattern Recognition}, pages 3977--3986, 2019.

\bibitem{zhang2020top}
Xiaoqin Zhang, Mingyu Fan, Di~Wang, Peng Zhou, and Dacheng Tao.
\newblock Top-k feature selection framework using robust 0--1 integer
  programming.
\newblock {\em IEEE Transactions on Neural Networks and Learning Systems},
  32(7):3005--3019, 2020.

\bibitem{bengio2020machine}
Yoshua Bengio, Andrea Lodi, and Antoine Prouvost.
\newblock Machine learning for combinatorial optimization: a methodological
  tour d’horizon.
\newblock {\em European Journal of Operational Research}, 2020.

\bibitem{khalil2016learning}
Elias Khalil, Pierre Le~Bodic, Le~Song, George Nemhauser, and Bistra Dilkina.
\newblock Learning to branch in mixed integer programming.
\newblock In {\em Proceedings of the AAAI Conference on Artificial
  Intelligence}, volume~30, 2016.

\bibitem{alvarez2017machine}
Alejandro~Marcos Alvarez, Quentin Louveaux, and Louis Wehenkel.
\newblock A machine learning-based approximation of strong branching.
\newblock {\em INFORMS Journal on Computing}, 29(1):185--195, 2017.

\bibitem{gasse2019exact}
Maxime Gasse, Didier Ch{\'e}telat, Nicola Ferroni, Laurent Charlin, and Andrea
  Lodi.
\newblock Exact combinatorial optimization with graph convolutional neural
  networks.
\newblock {\em arXiv preprint arXiv:1906.01629}, 2019.

\bibitem{gupta2020hybrid}
Prateek Gupta, Maxime Gasse, Elias~B Khalil, M~Pawan Kumar, Andrea Lodi, and
  Yoshua Bengio.
\newblock Hybrid models for learning to branch.
\newblock {\em arXiv preprint arXiv:2006.15212}, 2020.

\bibitem{tang2020reinforcement}
Yunhao Tang, Shipra Agrawal, and Yuri Faenza.
\newblock Reinforcement learning for integer programming: Learning to cut.
\newblock In {\em International Conference on Machine Learning}, pages
  9367--9376. PMLR, 2020.

\bibitem{vinyals2015pointer}
Oriol Vinyals, Meire Fortunato, and Navdeep Jaitly.
\newblock Pointer networks.
\newblock {\em arXiv preprint arXiv:1506.03134}, 2015.

\bibitem{bello2016neural}
Irwan Bello, Hieu Pham, Quoc~V Le, Mohammad Norouzi, and Samy Bengio.
\newblock Neural combinatorial optimization with reinforcement learning.
\newblock {\em arXiv preprint arXiv:1611.09940}, 2016.

\bibitem{kool2018attention}
Wouter Kool, Herke Van~Hoof, and Max Welling.
\newblock Attention, learn to solve routing problems!
\newblock {\em arXiv preprint arXiv:1803.08475}, 2018.

\bibitem{andrychowicz2016learning}
Marcin Andrychowicz, Misha Denil, Sergio Gomez, Matthew~W Hoffman, David Pfau,
  Tom Schaul, Brendan Shillingford, and Nando De~Freitas.
\newblock Learning to learn by gradient descent by gradient descent.
\newblock {\em arXiv preprint arXiv:1606.04474}, 2016.

\bibitem{li2016learning}
Ke~Li and Jitendra Malik.
\newblock Learning to optimize.
\newblock {\em arXiv preprint arXiv:1606.01885}, 2016.

\bibitem{nowak2017note}
Alex Nowak, Soledad Villar, Afonso~S Bandeira, and Joan Bruna.
\newblock A note on learning algorithms for quadratic assignment with graph
  neural networks.
\newblock {\em stat}, 1050:22, 2017.

\bibitem{hussein2017imitation}
Ahmed Hussein, Mohamed~Medhat Gaber, Eyad Elyan, and Chrisina Jayne.
\newblock Imitation learning: A survey of learning methods.
\newblock {\em ACM Computing Surveys (CSUR)}, 50(2):1--35, 2017.

\bibitem{torabi2019recent}
Faraz Torabi, Garrett Warnell, and Peter Stone.
\newblock Recent advances in imitation learning from observation.
\newblock {\em arXiv preprint arXiv:1905.13566}, 2019.

\bibitem{abbeel2004apprenticeship}
Pieter Abbeel and Andrew~Y Ng.
\newblock Apprenticeship learning via inverse reinforcement learning.
\newblock In {\em Proceedings of the twenty-first international conference on
  Machine learning}, page~1, 2004.

\bibitem{teerapittayanon2016branchynet}
Surat Teerapittayanon, Bradley McDanel, and Hsiang-Tsung Kung.
\newblock Branchynet: Fast inference via early exiting from deep neural
  networks.
\newblock In {\em 2016 23rd International Conference on Pattern Recognition
  (ICPR)}, pages 2464--2469. IEEE, 2016.

\bibitem{prechelt1998early}
Lutz Prechelt.
\newblock Early stopping-but when?
\newblock In {\em Neural Networks: Tricks of the trade}, pages 55--69.
  Springer, 1998.

\bibitem{yao2007early}
Yuan Yao, Lorenzo Rosasco, and Andrea Caponnetto.
\newblock On early stopping in gradient descent learning.
\newblock {\em Constructive Approximation}, 26(2):289--315, 2007.

\bibitem{kaya2019shallow}
Yigitcan Kaya, Sanghyun Hong, and Tudor Dumitras.
\newblock Shallow-deep networks: Understanding and mitigating network
  overthinking.
\newblock In {\em International Conference on Machine Learning}, pages
  3301--3310. PMLR, 2019.

\bibitem{shiryaev2007optimal}
Albert~N Shiryaev.
\newblock {\em Optimal stopping rules}, volume~8.
\newblock Springer Science \& Business Media, 2007.

\bibitem{becker2019deep}
Sebastian Becker, Patrick Cheridito, and Arnulf Jentzen.
\newblock Deep optimal stopping.
\newblock {\em Journal of Machine Learning Research}, 20:74, 2019.

\bibitem{chen2020learning}
Xinshi Chen, Hanjun Dai, Yu~Li, Xin Gao, and Le~Song.
\newblock Learning to stop while learning to predict.
\newblock In {\em International Conference on Machine Learning}, pages
  1520--1530. PMLR, 2020.

\bibitem{gintner2005solving}
Vitali Gintner, Natalia Kliewer, and Leena Suhl.
\newblock Solving large multiple-depot multiple-vehicle-type bus scheduling
  problems in practice.
\newblock {\em OR Spectrum}, 27(4):507--523, 2005.

\bibitem{helber2010fix}
Stefan Helber and Florian Sahling.
\newblock A fix-and-optimize approach for the multi-level capacitated lot
  sizing problem.
\newblock {\em International Journal of Production Economics}, 123(2):247--256,
  2010.

\bibitem{dorneles2014fix}
{\'A}rton~P Dorneles, Olinto~CB de~Ara{\'u}jo, and Luciana~S Buriol.
\newblock A fix-and-optimize heuristic for the high school timetabling problem.
\newblock {\em Computers \& Operations Research}, 52:29--38, 2014.

\bibitem{wang2011effective}
Yang Wang, Zhipeng L{\"u}, Fred Glover, and Jin-Kao Hao.
\newblock Effective variable fixing and scoring strategies for binary quadratic
  programming.
\newblock In {\em European Conference on Evolutionary Computation in
  Combinatorial Optimization}, pages 72--83. Springer, 2011.

\bibitem{gendreau2005tabu}
Michel Gendreau and Jean-Yves Potvin.
\newblock Tabu search.
\newblock In {\em Search methodologies}, pages 165--186. Springer, 2005.

\bibitem{chien2021hierarchical}
Jen-Tzung Chien and Chun-Wei Wang.
\newblock Hierarchical and self-attended sequence autoencoder.
\newblock {\em IEEE Transactions on Pattern Analysis and Machine Intelligence},
  2021.

\bibitem{he2016deep}
Kaiming He, Xiangyu Zhang, Shaoqing Ren, and Jian Sun.
\newblock Deep residual learning for image recognition.
\newblock In {\em Proceedings of the IEEE conference on computer vision and
  pattern recognition}, pages 770--778, 2016.

\bibitem{ioffe2015batch}
Sergey Ioffe and Christian Szegedy.
\newblock Batch normalization: Accelerating deep network training by reducing
  internal covariate shift.
\newblock In {\em International conference on machine learning}, pages
  448--456. PMLR, 2015.

\bibitem{achterberg2009hybrid}
Tobias Achterberg and Timo Berthold.
\newblock Hybrid branching.
\newblock In {\em International Conference on AI and OR Techniques in
  Constriant Programming for Combinatorial Optimization Problems}, pages
  309--311. Springer, 2009.

\bibitem{achterberg2009scip}
Tobias Achterberg.
\newblock Scip: solving constraint integer programs.
\newblock {\em Mathematical Programming Computation}, 1(1):1--41, 2009.

\bibitem{perez2018film}
Ethan Perez, Florian Strub, Harm De~Vries, Vincent Dumoulin, and Aaron
  Courville.
\newblock Film: Visual reasoning with a general conditioning layer.
\newblock In {\em Proceedings of the AAAI Conference on Artificial
  Intelligence}, volume~32, 2018.

\bibitem{boykov2004experimental}
Yuri Boykov and Vladimir Kolmogorov.
\newblock An experimental comparison of min-cut/max-flow algorithms for energy
  minimization in vision.
\newblock {\em IEEE transactions on pattern analysis and machine intelligence},
  26(9):1124--1137, 2004.

\bibitem{shi2000normalized}
Jianbo Shi and Jitendra Malik.
\newblock Normalized cuts and image segmentation.
\newblock {\em IEEE Transactions on pattern analysis and machine intelligence},
  22(8):888--905, 2000.

\bibitem{dantzig2016linear}
George Dantzig.
\newblock Linear programming and extensions.
\newblock In {\em Linear programming and extensions}. Princeton university
  press, 2016.

\bibitem{koller2009probabilistic}
Daphne Koller and Nir Friedman.
\newblock {\em Probabilistic graphical models: principles and techniques}.
\newblock MIT press, 2009.

\bibitem{everingham2015pascal}
Mark Everingham, SM~Eslami, Luc Van~Gool, Christopher~KI Williams, John Winn,
  and Andrew Zisserman.
\newblock The pascal visual object classes challenge: A retrospective.
\newblock {\em International journal of computer vision}, 111(1):98--136, 2015.

\bibitem{krizhevsky2009learning}
Alex Krizhevsky, Geoffrey Hinton, et~al.
\newblock Learning multiple layers of features from tiny images.
\newblock 2009.

\bibitem{deng2009imagenet}
Jia Deng, Wei Dong, Richard Socher, Li-Jia Li, Kai Li, and Li~Fei-Fei.
\newblock Imagenet: A large-scale hierarchical image database.
\newblock In {\em 2009 IEEE conference on computer vision and pattern
  recognition}, pages 248--255. Ieee, 2009.

\bibitem{carlini2017towards}
Nicholas Carlini and David Wagner.
\newblock Towards evaluating the robustness of neural networks.
\newblock In {\em 2017 ieee symposium on security and privacy (sp)}, pages
  39--57. IEEE, 2017.

\bibitem{szegedy2016rethinking}
Christian Szegedy, Vincent Vanhoucke, Sergey Ioffe, Jon Shlens, and Zbigniew
  Wojna.
\newblock Rethinking the inception architecture for computer vision.
\newblock In {\em Proceedings of the IEEE conference on computer vision and
  pattern recognition}, pages 2818--2826, 2016.

\bibitem{su2019one}
Jiawei Su, Danilo~Vasconcellos Vargas, and Kouichi Sakurai.
\newblock One pixel attack for fooling deep neural networks.
\newblock {\em IEEE Transactions on Evolutionary Computation}, 23(5):828--841,
  2019.

\bibitem{croce2019sparse}
Francesco Croce and Matthias Hein.
\newblock Sparse and imperceivable adversarial attacks.
\newblock In {\em Proceedings of the IEEE/CVF International Conference on
  Computer Vision}, pages 4724--4732, 2019.

\bibitem{modas2019sparsefool}
Apostolos Modas, Seyed-Mohsen Moosavi-Dezfooli, and Pascal Frossard.
\newblock Sparsefool: a few pixels make a big difference.
\newblock In {\em Proceedings of the IEEE/CVF conference on computer vision and
  pattern recognition}, pages 9087--9096, 2019.

\bibitem{xu2018structured}
Kaidi Xu, Sijia Liu, Pu~Zhao, Pin-Yu Chen, Huan Zhang, Quanfu Fan, Deniz
  Erdogmus, Yanzhi Wang, and Xue Lin.
\newblock Structured adversarial attack: Towards general implementation and
  better interpretability.
\newblock {\em arXiv preprint arXiv:1808.01664}, 2018.

\end{thebibliography}

\end{document}